# Trans-Fe elements from Type Ia Supernovae

## I. Heavy element nucleosynthesis during the formation of near-Chandrasekhar white dwarfs

U. Battino,[1,2,3] J. D. Keegans,[4,3] M. Allen,[4,5] F. K. Röpke,[6,7,8] F. Herwig,[9,10,3] A. Best,[1,11] R. Hirschi,[12,3] L. Piersanti,[2,13] O. Straniero,[2,13] S. A. Sim,[14] C. Travaglio,[15] and P. A. Denissenkov[9,10,3]

[1] Department of Physics, University of Naples Federico II, Via Cintia, Napoli, 80126, NA, Italy e-mail: umberto.battino@unina.it
[2] INAF – Osservatorio Astronomico d'Abruzzo, Via M. Maggini, 64100 Teramo, Italy e-mail: umberto.battino@inaf.it
[3] The NuGrid Collaboration ⋆
[4] E. A. Milne Centre for Astrophysics, University of Hull, Cottingham Road, Kingston upon Hull, HU6 7RX
[5] School of Mathematical and Physical Sciences, Hicks Building, Hounsfield Road, Sheffield, S3 7RH
[6] Heidelberg Institute for Theoretical Studies, Schloss-Wolfsbrunnenweg 35, 69118 Heidelberg, Germany
[7] Zentrum für Astronomie der Universität Heidelberg, Institut für Theoretische Astrophysik, Philosophenweg 12, 69120 Heidelberg, Germany
[8] Zentrum für Astronomie der Universität Heidelberg, Astronomisches Recheninstitut, Mönchhofstr. 12–14, 69120 Heidelberg, Germany
[9] Department of Physics & Astronomy, University of Victoria, Victoria, BC V8W 2Y2, Canada
[10] Joint Institute for Nuclear Astrophysics - Center for the Evolution of the Elements
[11] INFN – Sezione di Napoli, Via Cintia, 80126, Naples, Italy.
[12] School of Chemical and Physical Sciences, Keele University, Lennard-Jones Laboratories, Keele, ST5 5BG
[13] INFN – Sezione di Roma, Piazzale Aldo Moro 2, 00185, Rome, Italy.
[14] School of Mathematics and Physics, Queen's University Belfast, Belfast BT7 1NN, UK
[15] INAF – Osservatorio Astrofisico di Torino, Via Osservatorio, 20, 10025 Pino Torinese TO, Italy

**ABSTRACT**

*Context.* A Type Ia supernova (SNIa) marks the catastrophic explosion of a white dwarf in a binary system. These events play a crucial role in galactic chemical evolution and serve as pivotal standardizable candles for measuring cosmic distances, underpinning the discovery of the Universe's accelerated expansion. However, the progenitors of SNIa remain uncertain, introducing challenges to their use in cosmology and nucleosynthesis predictions.
*Aims.* In this work, we present a grid of five models detailing the evolution and nucleosynthesis of slowly merging carbon-oxygen white dwarfs approaching the Chandrasekhar mass.
*Methods.* These models test a variety of physics input settings, including accretion rates, nuclear reaction rates, convection parameters, and the composition of the accreted material. During the merger process, as the mass of the primary white dwarf approaches the Chandrasekhar limit, carbon burning is initiated first on the surface before eventually igniting explosively at the center. As a consequence, the $^{22}$Ne($\alpha$,n)$^{25}$Mg reaction activates in the outer layers of all models.
*Results.* The neutrons released in this way produce a weak *s*-process-like abundance distribution peaking at Kr, which is overproduced by more than a factor of $\sim$1000 compared to solar. The trans-Fe elements-enriched outer layer mass varies from $\sim$0.04 $M_\odot$ to $\sim$0.11 $M_\odot$, depending on the accretion rate. Our explosion simulation of these progenitor models ejects significant amount of first-peak elements (e.g., Kr, Sr) as well as of some long-lived radioactive species, such as $^{60}$Fe.
*Conclusions.* In a previous theoretical study, we found that a similar nucleosynthesis process during the progenitor phase may also occur on the surface of near-Chandrasekhar white dwarfs formed through the accretion of H-rich material via the single-degenerate scenario. Therefore, these results suggest trans-Fe enrichment might be a hallmark of near-Chandrasekhar SNIa ejecta, regardless of the specific progenitor channel, and could provide a new spectral signature distinguishing them from sub-Chandrasekhar explosions.

## 1. Introduction

It is known that about half of low- and intermediate-mass stars (M < 8 $M_\odot$) in the Solar Neighborhood have at least one companion (Moe & Di Stefano 2017). At the end of their evolution, these stars lose their entire envelope (Iben & Renzini 1983; Herwig 2005), leaving behind their bare, compact carbon-oxygen (CO) core, where no nuclear reactions occur. The resulting structure is supported solely by electron degeneracy pressure and is known as a white dwarf (WD). If a WD is in a binary system, it can accrete matter from a companion, potentially triggering a type Ia supernova (SNIa) explosion. Two main types of explosions are possible: near-Chandrasekhar and sub-Chandrasekhar. Both scenarios can arise from two primary evolutionary channels: the single-degenerate channel (Whelan & Iben 1973; Nomoto et al. 2007) and the double-degenerate channel (Iben 1984; Pakmor et al. 2011). Near-Chandrasekhar SNIa from the single-degenerate channel originate from a WD that accretes mass from the H-rich envelope of a less evolved stellar companion. In contrast, near-Chandrasekhar SNIa from the double-degenerate channel involve a companion that is another WD, with a mass ratio of the components less than $\sim 0.8$ (Pakmor et al. 2011). In this case, the separation between the two WDs decreases as a consequence of gravitational waves emission until they merge; the less massive (i.e., secondary) WD is tidally disrupted, forming a CO-rich accretion disk around the primary WD. In both cases, a slow accretion process ensues, and the primary WD gradually approaches the critical Chandrasekhar mass (the maximum mass that can be supported by electron degeneracy, around 1.39 $M_{\odot}$ for a CO WD). Carbon burning then ignites explosively, causing a SNIa. In the case of sub-Chandrasekhar SNIa from the single-degenerate channel, the accreted material is mainly helium, which can trigger a detonation before the Chandrasekhar mass is reached (see e.g. Livne 1990; Woosley & Weaver 1994; Piersanti et al. 2014). Finally, sub-Chandrasekhar SNIa from the double-degenerate channel result from the violent merger of two WDs of similar mass (Pakmor et al. 2012), or if a He shell thick enough (at least $\sim 0.01$ $M_{\odot}$) is present on top of both WDs (Pakmor et al. 2013; Shen et al. 2024; Pakmor et al. 2022).

⋆ http://www.nugridstars.org

In the standard Big Bang Nucleosynthesis picture, apart from H, He, and Li, every chemical element in the Universe was first forged by sequences of nuclear reactions (called nucleosynthesis processes) in stars and related explosions (Burbidge et al. 1957; Cameron 1957). In this way, the composition of our Milky Way, as well as that of all other galaxies, progressively changed from a primordial one (consisting of only H, He, and Li) to one increasingly enriched in heavier elements. This process is known as galactic chemical evolution (GCE). SNIa are the major producers of Fe-peak elements and highly relevant also for the α-element content in the Galaxy (Matteucci & Greggio 1986; Kobayashi et al. 2020). Additionally, the potential relevance of SNIa in the origin of other elements depends on the fraction of SNIa produced via each of the aforementioned channels. For example, Travaglio et al. (2015) (hereafter TRV15) showed that near-Chandrasekhar single-degenerate SNIa can be a significant source of a class of trans-Fe, proton-rich isotopes known as *p*-nuclei (produced by the so-called *p*-process (Arnould & Goriely 2003)). The main assumption of TRV15, later confirmed by Battino et al. (2020), was the crucial presence of a trans-Fe neutron-capture element distribution up to Pb (referred to as the "seed distribution") formed on the WD surface during the accretion of H-rich material from the companion. Photon-induced reactions on this seed distribution during the explosion can then result in the formation of *p*-nuclei.

Quantifying the fraction of each SNIa progenitor scenario is therefore a matter of urgency, both for the role of SNIa in GCE and in cosmology. In particular, the relative contributions of near-Chandrasekhar and sub-Chandrasekhar explosions are still not fully understood. Johansson et al. (2016), based on spectra from a large sample of early-type galaxies, found that near-Chandrasekhar single-degenerate SNIa should account for less than 6% of the total SNIa population. This is consistent with WD models accreting H-rich material at solar metallicity from both Denissenkov et al. (2017) and Battino et al. (2020). Denissenkov et al. (2017) showed that the low mass-retention efficiency following He flashes, naturally arising during accretion, would severely limit the single-degenerate path to near-Chandrasekhar masses. Battino et al. (2020) estimated the absolute minimum initial mass a WD must have in order to reach the Chandrasekhar limit via accretion of H-rich material at around solar metallicity to be $\sim 0.91$ $M_{\odot}$. Considering that O'Brien et al. (2024) found $\sim 3\%$ of white dwarfs in their 40 pc sample with masses $>1$ $M_{\odot}$, and about the same fraction between $\sim 0.9$ $M_{\odot}$ and $\sim 1$ $M_{\odot}$, this corresponds to $\sim 6\%$ of the local white dwarf population. On the other hand, Lach et al. (2020) computed the nucleosynthetic ejecta of sub-Chandrasekhar single-degenerate SNIa and re-determined the contributions of both near- and sub-Chandrasekhar events to the total SNIa rate by simultaneously reproducing the observed [Ti/Fe], [Mn/Fe], and [V/Fe] in the Galaxy. They found that at least 45% of all SNIa should originate from near-Chandrasekhar mass explosions, as a too large number of sub-Chandrasekhar SNIa from He detonations would lead to tensions in [Ti/Fe] and [V/Fe] since these elements are produced in super-solar amounts in them. This result was more recently qualitatively confirmed by further studies. Shen et al. (2024) showed that most WDs would be born with enough He for its ignition to be plausible, implying that the majority of SNIa could originate from sub-Chandrasekhar explosions. However, aside from the uncertainties introduced by their 2D treatment of inherently 3D processes leading to detonation, the Fe masses produced in the models of Shen et al. (2024) (once $^{56}$Ni has fully decayed) remain relatively low, even accounting for the early compression of the secondary WD by the primary detonation (at most $\sim 0.15$ $M_{\odot}$ in the case of a primary WD mass $<1$ $M_{\odot}$, representative of the majority of CO WDs, see also Boos et al. 2024). This may confirm that He shell detonation sub-Chandrasekhar SNIa, while still very plausibly responsible for the majority of SNIa, cannot fully replace near-Chandrasekhar SNIa without resulting in super-solar [Ti/Fe] and [V/Fe] values. Moreover, Trueman et al. (2025) attempted to calculate the sub-Chandrasekhar SNIa fractions ($f_{sub}$) needed to fit the evolution of one Fe-peak element at a time versus [Fe/H]. While their $f_{sub}$ varied substantially depending on which tracer element was considered, a minimum near-Chandrasekhar fraction of around 40% appears to be a reasonable average of their results. However, this conclusion depends on the massive star ejecta yields adopted in the calculations (see also Gronow et al. 2021). Therefore, if the contribution of near-Chandrasekhar single-degenerate SNIa to the total rate is observationally constrained to be below $\sim 6\%$, then the possibility that $\sim 34\%$ of near-Chandrasekhar SNIa originate from the double-degenerate channel must be considered. In principle, this may occur not only through the direct accretion of CO-rich material from a CO WD, but also via He from a He star (Piersanti et al. 2014). However, this channel may be severely limited by the simultaneous requirement of both a population of significantly massive He stars ($M \geq 0.8$ $M_{\odot}$) with massive companions ($M_{WD} \geq 1.0$ $M_{\odot}$), and a very specific accretion rate (between $1\times10^{-6}$ $M_{\odot}$ yr$^{-1}$ and $2\times10^{-6}$ $M_{\odot}$ yr$^{-1}$). Higher accretion rates would lead to the formation of an ONeMg WD, thus not producing a SNIa, while lower rates would trigger strong He flashes once the primary WD exceeds $\sim 1.2$ $M_{\odot}$, making further accretion difficult due to the resulting high mass loss. The limitations of direct He accretion up to the Chandrasekhar mass are therefore somewhat similar to those affecting H accretion via the single-degenerate scenario, and may likewise contribute to only a small fraction of SNIa. This leaves a residual $\sim 30\%$ that must originate from an alternative source, such as slow CO WD mergers. This remains an open question, with three main shortcomings identified. First, during disk formation, C could ignite near the accreting star's surface and propagate rapidly inward, forming an ONeMg WD instead of causing a thermonuclear ex-plosion. This was ruled out by Lorén-Aguilar et al. (2009), who found that although C burning does occur, it is quickly extinguished. Second, mass transfer from the disk to the primary WD could proceed at a high rate ($\sim 10^{-5}$ $M_{\odot}$ yr$^{-1}$), potentially causing an off-centre ignition of C burning before the WD reaches the Chandrasekhar mass. Such burning would then propagate toward the center, again possibly turning the whole structure into an ONeMg WD Siess (2006). However, Denissenkov et al. (2013) showed that even a small amount of convective boundary mixing removes the physical conditions required for the C-flame to propagate all the way to the center. Furthermore, Piersanti et al. (2003a) argued that rotation, naturally arising in merging WDs, can self-regulate the accretion rate to as low as a few $10^{-7}$ $M_{\odot}$ yr$^{-1}$. This last result was anyway contested by both Saio & Nomoto (2004) and Yoon & Langer (2005), who raised questions about the validity of the assumption of rigid rotation as well as of the high estimated ratio of rotational to gravitational energy. The third and final shortcoming is that the number of observationally confirmed double-degenerate systems near the Chandrasekhar mass limit is very small ($<$5 systems; see Breedt et al. (2017)). However, population synthesis estimates suggest the average WD merger timescale in the Galaxy is $\sim$0.7–1 Gyr (Toonen et al. 2012). Using the observed Milky Way SNIa rate, roughly $4.6\times10^{6}$ SNIa would occur over this time. With $\sim 10^{9}$ WDs in the Galaxy (Harris et al. 2006), we would expect about one in $\sim$200 observed WDs to be a SNIa progenitor if SNIa arise solely from double-degenerate binaries. Allowing for a factor of $\sim$2 systematic uncertainty in the observed SNIa rate (Li et al. 2011) and not perfect detection efficiency in follow-up observations, the true number could easily be half this estimate or less (Breedt et al. 2017). Therefore, we should expect to find around 2 double-degenerate systems near the Chandrasekhar mass in every $\sim$1000 observed WDs. Interestingly, this is exactly what the SN Ia Progenitor Survey (SPY24) found: among 1014 catalogued WDs, it discovered 39 double-degenerate binaries, two of which had a total mass close to the Chandrasekhar limit. Furthermore, Cheng et al. (2020) searched for evidence of double-WD merger products by analyzing high-mass WDs in GAIA DR2. They obtained a precise double-WD merger rate consistent with binary population synthesis results and in agreement with SPY, supporting the idea that double-WD mergers may contribute significantly to the SNIa population.

In this work, we assume that Chandrasekhar-mass WDs can indeed form via the double-degenerate channel through a slow merger process mediated by an accretion disk. We compute the stellar evolution and nucleosynthesis during the progenitor phase i.e., the evolutionary phase of the system prior to its explosion as a supernova, beginning with the onset of mass accretion and ending with the ignition of explosive carbon burning at the centre) and investigate whether a trans-Fe element distribution can form in a manner similar to that observed in near-Chandrasekhar single-degenerate SNIa progenitors, as previously verified by Battino et al. (2020). The only study to have considered this hypothesis was presented by Battino et al. (2022), albeit in a very preliminary form. That study focused on a single composition for both the primary and secondary WDs, a single accretion rate, and only on the progenitor phase, no explosion was simulated. In this paper, we therefore aim to expand on the initial study by Battino et al. (2022) by exploring a range of accretion rates and CO-rich compositions accreted onto the primary WD. Additionally, we will compute the nucleosynthesis resulting from the explosion of our progenitor models to assess whether significant amounts of trans-Fe elements are ejected. If so, the potential spectral identification of trans-Fe elements in "normal" Type Ia supernova emission could serve as a new signature of the explosion of a near-Chandrasekhar mass WD. No sub-Chandrasekhar explosive model in the literature is known to eject substantial amounts of first-peak elements (such as Kr, Rb, Sr, Y, Zr) or heavier species. Woosley & Kasen (2011) reported overproduction factors for some Kr isotopes, but values exceeding 100 occur only for the p-only nucleus $^{78}$Kr, implying that the overall elemental overproduction compared to solar, and the total ejected Kr mass, remains relatively low, on the order of a few $10^{-8}$M$_{\odot}$. Keegans et al. (2023) computed complete abundance yields using a full nuclear reaction network up to Bi for 39 models of SNIa explosions, based on three progenitors: a 1.4M$\odot$ deflagration-detonation model, a 1.0M$_{\odot}$ double-detonation model, and a 0.8M$\odot$ double-detonation model. These were calculated for 13 metallicities, with $^{22}$Ne (i.e., the main neutron source during the explosion and critical for trans-Fe element nucleosynthesis) mass fractions ranging from 0 to the very high super-solar value of 0.1. Even in the models with the highest $^{22}$Ne abundances, the heaviest trans-Fe species showing any positive overproduction factor relative to solar was $^{70}$Ge, while no first-peak elements were produced. This is linked to the very short timescales between the onset of mass transfer and the final explosion in violent CO WD mergers (less than $\sim$10 minutes, Pakmor et al. 2012), and between the start of surface nuclear burning and the trigger of the He detonation in double-detonation models ($\sim$5 minutes, see Pakmor et al. 2013).

In the following pages, the input physics, relevant setups and codes adopted are described in Section 2. The calculation of full stellar and nucleosynthesis models during the progenitor phase are discussed in Sections 3 and 4 respectively, while our explosive nucleosynthesis calculations are presented in Section 5. Our conclusions are presented in Section 6.

## 2. Input physics and computational methods

The SNIa progenitor models presented in this work are computed using the 1D stellar code MESA (revision 10108). The CO WD model 0.927-from-6.0-z2m2.mod, included in the MESA data folder and with a mass of 0.927 $M_{\odot}$, was cooled until the effective temperature dropped below $3.5\times10^{5}$ K before being used as the starting model for all our progenitor calculations. The choice of this particular mass and cooling criterion was intended to facilitate comparison with the results of Wu et al. (2019), plotted in their Figure 2.

We employed three different sets of chemical compositions for the material accreted from the secondary WD, as shown in Table 1. The first set (Abu Set1) was obtained by simulating the evolution of a 1.65 $M_{\odot}$, Z = 0.01[1] star from the pre-main sequence up to the formation of the CO WD, using the same physics inputs as in Ritter et al. (2018). At the end of the evolution, the H-free core (defined as the mass enclosed within the region where the hydrogen mass fraction drops below $10^{-4}$) was fully mixed, and the resulting abundances were adopted as the accreted composition. The second and third sets (Abu Set2 and Abu Set3) were employed with the main goal of exploring the impact of changing the accreted abundances of three key species.

[1] Notice that this introduces an inconsistency with the primary WD originating from a Z=0.02 progenitor. This might influence the C and O profiles in the deep WD interior, potentially impacting the explosion conditions. However, this does not affect the results of the present work, which is mainly focused on the trans-Fe element production occurring in the outermost $\sim$0.1 $M_{\odot}$ when the WD mass has reached a near-Chandrasekhar value. As a consequence, this production depends solely on the accreted composition and not on the central conditions.

In particular, Abu Set2 is identical to Abu Set1, but with an enhanced abundance of the key neutron source $^{22}$Ne, which is qualitatively equivalent to accreting higher metallicity material. Abu Set3 builds on Abu Set2, but with equal mass fractions of C and O, both set to 0.481. A very important aspect is the presence of He in the accreted composition. Pakmor et al. (2013) simulated the merger of two CO WDs, each with a thin He shell (0.01 $M_\odot$) on top, and showed that the accretion of He onto the primary WD leads to the formation of a detonation in the He shell. This detonation can propagate inward and ignite explosive C-burning, resulting in a sub-Chandrasekhar SNIa explosion. However, in this work we assume that accretion occurs through an accretion disk that is both fully mixed and formed over a short timescale (typically in ∼1.5 minutes after the matter flowing out of the secondary hits the surface of the primary white dwarf for systems with a mass ratio ∼0.75, see Lorén-Aguilar et al. 2009). We therefore focus on the case where the explosion does not occur immediately during the early dynamical phase of the merger, but rather proceeds via disruption of the secondary and subsequent accretion from a disk. Furthermore, it is worth noting that the typical mass of the He intershell in a 6 $M_\odot$ star at approximately solar metallicity (i.e., the progenitor of the ∼0.9 $M_\odot$ WD considered in this study) is ∼0.004 $M_\odot$, with a typical He mass fraction of ∼0.5 (Ritter et al. 2018), resulting in ∼0.002 $M_\odot$ of He. This is roughly a factor of five lower than the 0.01 $M_\odot$ He shell considered in Pakmor et al. (2013). However, more work is needed to systematically explore the chances to ignite a He-detonation with lower He shell masses than that considered by Pakmor et al. (2013). This is indeed an important topic, particularly since Shen et al. (2024) recently showed that, assuming He can be ignited in these conditions, the propagation of a double detonation remains plausible even for such small He shell masses on the order of ∼0.002 $M_\odot$.

We used two different nuclear reaction networks during two distinct phases of the merging process. At the beginning of the mass increase, we employed the co-burn-extras.net MESA network, which primarily includes the isotopes required for C-burning up to $^{28}$Si, connected through more than 50 reactions. We extended the isotope list of co-burn-extras.net by adding the Fe-peak isotopes listed in Table 1, which act as seeds for neutron captures once the primary WD approaches the Chandrasekhar mass and C-burning ignites at the surface. Until this point, reactions beyond $^{28}$Si are negligible and were therefore excluded. When the temperature at the base of the accreted material exceeds 1 GK, C-burning is triggered, and the $^{22}$Ne($\alpha$,n)$^{25}$Mg neutron source becomes active in the outer layers of the primary WD. In this final stage, we extended the nuclear reaction network up to $^{157}$Eu, including a total of 429 isotopes from H to Eu. All isotopes in the network were connected through the relevant nuclear reactions in order to accurately follow the neutron-capture nucleosynthesis during surface C-burning.

Charged-particle induced thermonuclear reaction rates involving nuclei in the range $1 \leq Z \leq 14$ from the NACRE compilation (Angulo et al. 1999) were adopted if available. For all other reactions, up to $^{157}$Eu, the rates recommended in the JINA REACLIB library ReaclibV2.2 were adopted (see Cyburt et al. (2010)), except for the $^{12}$C+$^{12}$C reaction rate from Gasques et al. (2005), the $^{12}$C($\alpha$,$\gamma$)$^{16}$O from Kunz et al. (2002), and the $^{22}$Ne($\alpha$,n)$^{25}$Mg and $^{22}$Ne($\alpha$,$\gamma$)$^{26}$Mg from Adsley et al. (2021). Weak reactions rates are mainly based on the tabulations of Langanke & Martínez-Pinedo (2000), complemented by Fuller et al. (1985) and Oda et al. (1994).

Convective mixing was modeled using the standard mixing length theory (MLT, Böhm-Vitense 1958), in which the free mixing-length parameter $\alpha_{\rm MLT}$ defines the mean free path (i.e., the mixing length) of a convective element in units of the pressure scale height. Convective boundary mixing (CBM) was also included, following the exponential overshooting formalism of Herwig (2000):

$$D(dr) = D_0 \times \exp(-2dr/f_1 Hp_0) \quad (1)$$

where dr is the geometric distance to the convective boundary. The term $f_1 \times Hp_0$ identifies the scale height of the overshooting regime. The values $D_0$ and $Hp_0$ are respectively the diffusion coefficient and the pressure scale height at the convective boundary. $f_1$ is set to 0.004 under H-shells (as in Denissenkov et al. (2014)) and to 0.01 under He- (as found by Herwig et al. (2007)) and heavier shells.

The CO-enhanced (Type 2) opacities were used throughout the calculations, based on OPAL tables (Iglesias & Rogers 1996). To stabilize the initial phase, just before the onset of net mass increase, dominated by thermal instabilities, we reduced the opacity to suppress the iron opacity bump by applying an opacity reduction factor of 0.01 in the temperature range $5.3 < \log T < 5.7$. Electron conduction opacities are from Cassisi et al. (2007). In case density or temperature conditions are outside the region covered by Cassisi et al. (2007), the Iben (1975) fit to the Hubbard & Lampe (1969) electron conduction opacity is used for non-degenerate conditions, while the Yakovlev & Urpin (1980) fits are used for degenerate ones.

The main settings adopted in the grid of SNIa progenitor models presented in this work are shown in Table 2. To explore the impact of different physical inputs on the nucleosynthesis, each model is characterised by a specific combination of accreted abundances, mass accretion rate, and the value of the $\alpha_{\rm MLT}$ parameter. In particular, we focus on accretion rates > $2\times10^{-6}$ $M_\odot$yr$^{-1}$, as these are the conditions necessary for surface C-burning ignition (Piersanti et al. 2003b) and, therefore, for the production of trans-Fe elements, the main focus of the present work.

To compute the nucleosynthesis resulting from the explosion of our progenitors, we used the NuGrid code tppnp (tracer particle post-processing network-parallel). A full description of the tppnp code can be found in Jones et al. (2019), with additional details about the nuclear reaction network provided in Ritter et al. (2018) and Pignatari et al. (2016). Additional details about the tppnp code are given in Section 5.

## 3. Progenitor structure models

### 3.1. General features

The HR diagrams of the models introduced in Table 2 are shown in Figure 1. The tracks of models A, B, and C are very similar, while significant differences arise when comparing them to models D and E. Due to the increased $^{12}$C abundance in the accreted material, surface C-burning in Model E ignites at a lower temperature compared to models A, B, and C. This causes the evolutionary track to shift towards lower temperatures as C-burning ignites and increases the luminosity by more than an order of magnitude. Similarly, the higher accretion rate in Model D causes the temperature at the base of the accreted material to rise more rapidly, reaching surface C-burning conditions earlier. As a result, the mass of the primary WD at the time of ignition is lower, and consequently, the degree of electron degeneracy in the WD's plasma is also reduced. For this reason, surface C-ignition occurs at a lower temperature compared to all other

Table 1: Sets of chemical composition abundances in mass fraction of the accreted material from the secondary WD used in this work.

| Isotope | Abu Set1 | Abu Set2 | Abu Set3 |
|---|---|---|---|
| $^{1}$H | 7.125E-07 | 7.125E-07 | 7.125E-07 |
| $^{3}$He | 6.899E-19 | 6.899E-19 | 6.899E-19 |
| $^{4}$He | 0.016E+00 | 0.016E+00 | 0.016E+00 |
| $^{7}$Li | 5.346E-17 | 5.346E-17 | 5.346E-17 |
| $^{7}$Be | 6.170E-19 | 6.170E-19 | 6.170E-19 |
| $^{8}$B | 8.802E-23 | 8.802E-23 | 8.802E-23 |
| $^{12}$C | 0.365E+00 | 0.365E+00 | 0.481E+00 |
| $^{13}$C | 1.567E-07 | 1.567E-07 | 1.567E-07 |
| $^{14}$N | 1.493E-04 | 1.493E-04 | 1.493E-04 |
| $^{15}$N | 1.418E-07 | 1.418E-07 | 1.418E-07 |
| $^{16}$O | 0.602E+00 | 0.597E+00 | 0.481E+00 |
| $^{17}$O | 1.193E-06 | 1.193E-06 | 1.193E-06 |
| $^{18}$O | 4.899E-10 | 4.899E-10 | 4.899E-10 |
| $^{19}$F | 3.451E-10 | 3.451E-10 | 3.451E-10 |
| $^{20}$Ne | 0.002E+00 | 0.002E+00 | 0.002E+00 |
| $^{22}$Ne | 0.011E+00 | 0.016E+00 | 0.016E+00 |
| $^{23}$Na | 1.075E-04 | 1.075E-04 | 1.075E-04 |
| $^{25}$Mg | 3.867E-05 | 3.867E-05 | 3.867E-05 |
| $^{26}$Mg | 0.003E+00 | 0.003E+00 | 0.003E+00 |
| $^{50}$Cr | 4.000E-07 | 4.000E-07 | 4.000E-07 |
| $^{52}$Cr | 8.000E-06 | 8.000E-06 | 8.000E-06 |
| $^{53}$Cr | 1.000E-06 | 1.000E-06 | 1.000E-06 |
| $^{54}$Cr | 2.000E-07 | 2.000E-07 | 2.000E-07 |
| $^{55}$Mn | 5.000E-06 | 5.000E-06 | 5.000E-06 |
| $^{54}$Fe | 4.000E-05 | 4.000E-05 | 4.000E-05 |
| $^{56}$Fe | 7.000E-04 | 7.000E-04 | 7.000E-04 |
| $^{57}$Fe | 2.000E-05 | 2.000E-05 | 2.000E-05 |
| $^{59}$Co | 2.000E-06 | 2.000E-06 | 2.000E-06 |
| $^{58}$Ni | 3.000E-05 | 3.000E-05 | 3.000E-05 |
| $^{60}$Ni | 1.000E-05 | 1.000E-05 | 1.000E-05 |
| $^{61}$Ni | 5.000E-07 | 5.000E-07 | 5.000E-07 |
| $^{62}$Ni | 2.000E-06 | 2.000E-06 | 2.000E-06 |
| $^{64}$Ni | 4.000E-07 | 4.000E-07 | 4.000E-07 |

Table 2: Main settings adopted in the grid of SNIa progenitor models presented in this work.

| Model | Acc Abundance | Acc Rate [$M_\odot$ yr$^{-1}$] | $\alpha_{MLT}$ |
|---|---|---|---|
| Model A | Abu Set1 | 2.4E-06 | 2 |
| Model B | Abu Set2 | 2.4E-06 | 2 |
| Model C | Abu Set2 | 2.4E-06 | 1.6 |
| Model D | Abu Set2 | 4.3E-06 | 1.6 |
| Model E | Abu Set3 | 2.4E-06 | 1.6 |

**Notes.** The initial mass of the primary WD is 0.927 $M_\odot$ and it is the same for all models. For each model, the selected accredted abundance, the mass accretion rate and the adopted mixing length $\alpha_{MLT}$ parameter are specified. The accreted abundances refer to those detailed in table 1.

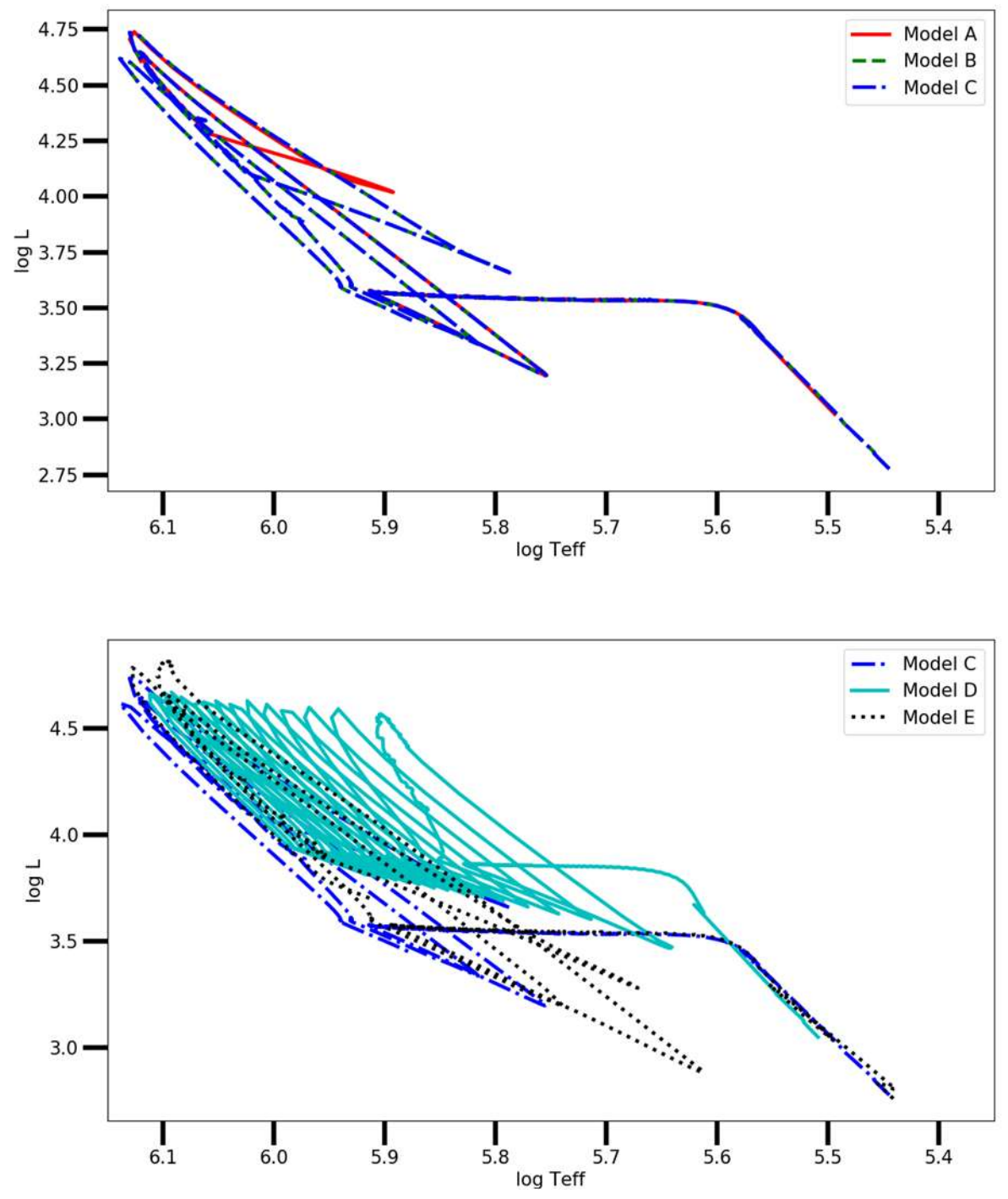

Fig. 1: HR diagrams of the models listed in table 1. The evolution starts at the bottom right end of the tracks.

models with lower accretion rates. This affects the HR track (as shown in the bottom panel of Figure 1), which is visibly shifted towards lower temperatures than any of the other models. Furthermore, in Model D, C-ignition proceeds as a series of flashes occurring at progressively higher temperatures as mass accretion continues.

These flashes are also visible in Figure 2, which shows the Kippenhahn diagrams of models D and C. In both models, the initial mass accretion proceeds steadily until the ignition of surface C-burning at a higher WD mass, the specific value of which depends on the mass accretion rate, and occurs before the onset of core C-burning that drives the simmering phase prior to the final explosion. In particular, in Model D, surface C-burning ignites at $M_{WD} \simeq 1.275\ M_\odot$. This is followed by 18 C-flashes, each driving the formation of a small convective region right at the WD surface. On the other hand, the lower accretion rate in Model C causes surface C-ignition to occur at a higher WD mass, $M_{WD} \simeq 1.374\ M_\odot$. As shown in Figure 3, this is primarily due to the higher temperature at the base of the accreted material, at a given WD mass, in models with a higher mass accretion rate. The temperature and density profiles of models C and D are presented at different stages during the mass accretion phase, up to the ignition of surface C-burning. This ignition is marked by the profile featuring a sharp temperature peak exceeding $10^9$ K at the base of the accreted material. These profiles closely resemble those obtained by Wu et al. (2019). At any point during the mass accretion, the temperature peak at the base of the accreted layer is determined by two main processes: 1) Thermal diffusion into the WD interior, which tends to lower the temperature, and 2) Compressional heating due to accreted mass, which tends to increase it (Piersanti et al. 2014). The higher mass deposition rate onto the primary WD in Model D results in a shorter compressional heating timescale at the accretor's surface. More specifically, the disparity between the compressional heating and the inward thermal diffusion timescales increases. Consequently, the temperature at the base of the accreted layer increases more rapidly and reaches a higher value in Model D than in Model C for the same primary WD mass during accretion. Ultimately, C-burning is ignited when the local temperature reaches approximately $10^9$ K.

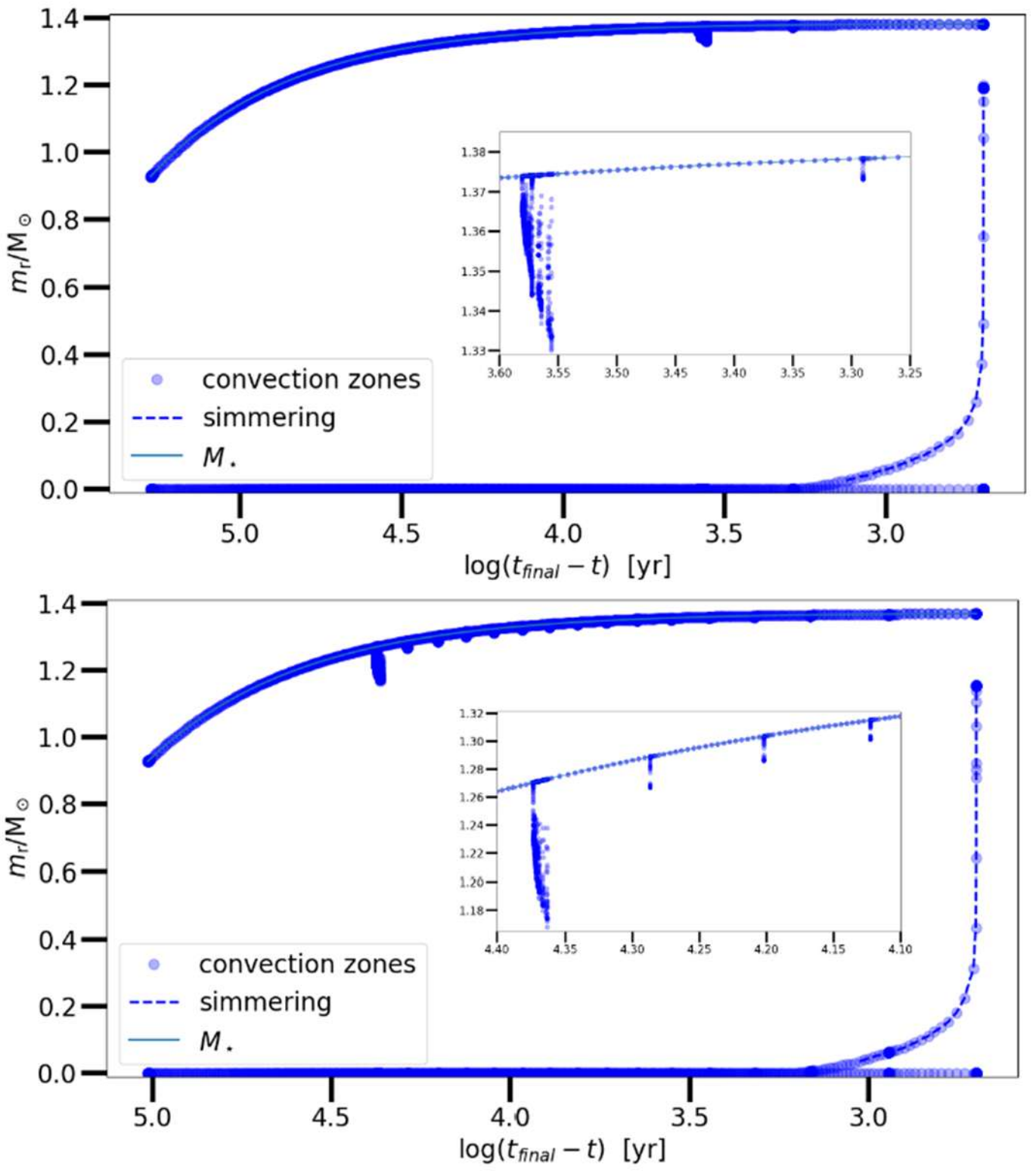

Fig. 2: Kippenhahn diagrams of Model C (top panel) and D (bottom panel)

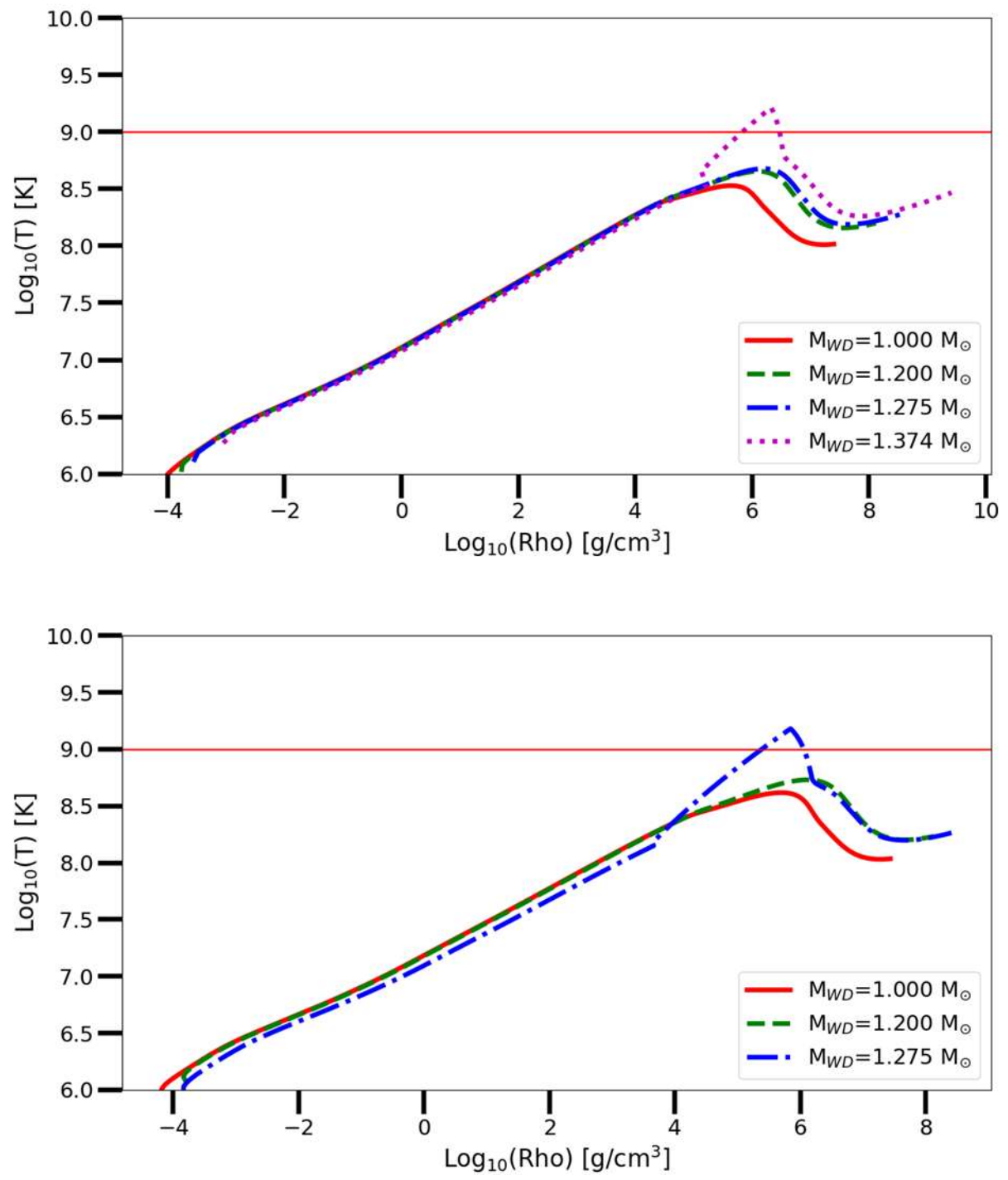

Fig. 3: Temperature and density profiles of Model C (upper panel) and Model D (lower panel) at different moments during the mass accretion.

### 3.2. C-flame propagation

Table 3 shows that each first surface C-flash lasts between 150 and 550 years, depending on the model. In particular, we note a significant difference (approximately 65%) between models B and C, where the only difference in the physical input is the value of the mixing length parameter α. This is illustrated in Figure 4, which presents the first surface flash in both models B and C through two Kippenhahn diagrams as a function of time. While the extent of the convective zone reaches the same mass coordinate in both models ($Mr$ = 1.33 $M_\odot$), it is reached over visibly different timescales, indicating a different propagation speed of the burning front. Indeed, in all stellar models discussed in this work, the first surface C-burning episode develops a C-flame front moving inward, inducing the formation of a convective shell that is dragged toward the stellar interior as the burning front advances. Figure 5 shows how the convective region induced by C-burning is tightly bound to the location of the temperature peak, corresponding to the position of the inward-moving burning front, and does not extend up to the surface once flame propagation begins. This implies that each zone will experience convective C-burning for a duration inversely proportional to the C-flame propagation speed, which will impact the final neutron exposure and the trans-Fe element distribution in a given zone.

In 1D stellar models, the impact of the mixing length parameter α on the C-flame propagation speed is linked to how the diffusion coefficient is defined within convectively unstable regions, i.e., $D_c = \frac{1}{3} v_c l$, where $l$ is the mixing length and $v_c$ is the average velocity of the convective elements according to the mixing length theory (MLT) (Böhm-Vitense 1958). Then the diffusion coefficient can be written as

$$D_c = \frac{1}{3}\alpha^{2/3}H_p\left(\frac{c}{\kappa\rho}g\beta(1-\beta)\nabla_{\rm ad}(\nabla_{\rm rad}-\nabla_{\rm ad})\right)^{1/3} \qquad (2)$$

where $H_p$ is the pressure scale height, $\kappa$ the opacity, $c$ the speed of light, $\rho$ the density, $g$ the gravitational acceleration, $\nabla_{rad}$ and $\nabla_{ad}$ the radiative and adiabatic temperature gradient respectively, and $\beta$ the gas pressure fraction. This leads to $D_c$, as well as $D_0$ in the CBM equation (2), being proportional to $\alpha^{2/3}$.

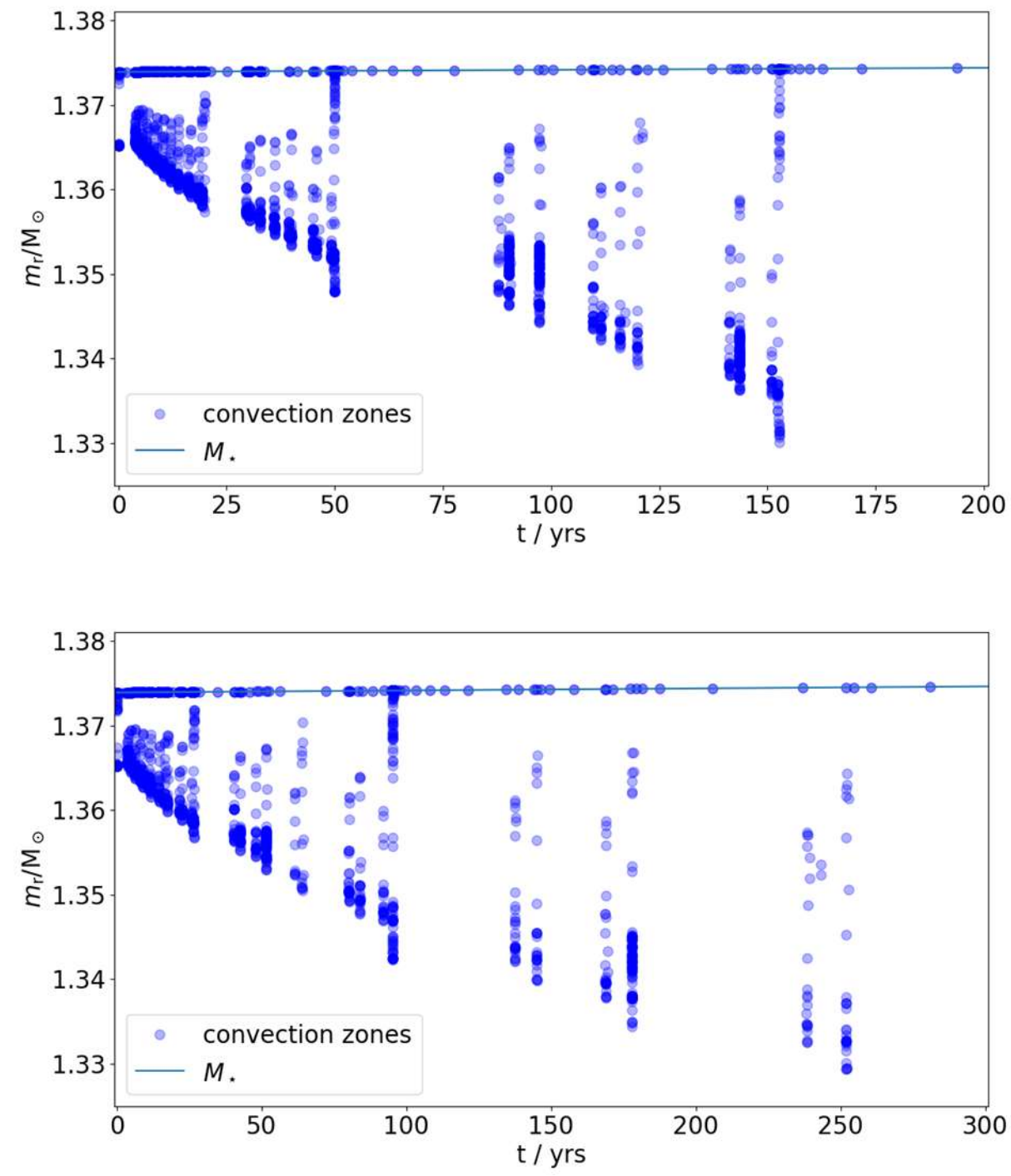

Fig. 4: Kippenhan diagram of the first surface C-flash and related propagation of C-flames in models B (top panel) and C (bottom panel). The X-axis shows the time from the onset of surface C-burning.

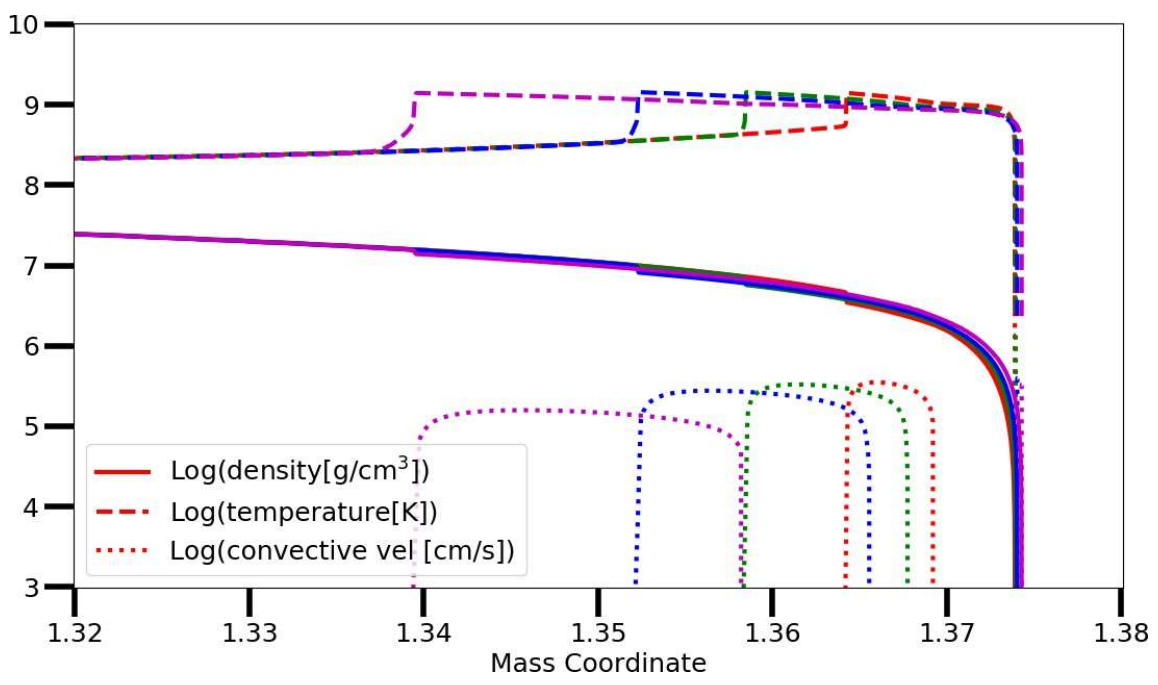

Fig. 5: Temperature, density and convective velocity profiles of Model C at different times during the development of the first surface C-burning episode. Lines of the same color correspond to the same simulation timestep. The convective region induced by C-burning is tightly bound to the location of the temperature peak, corresponding to the location of the inward-propagating burning front.

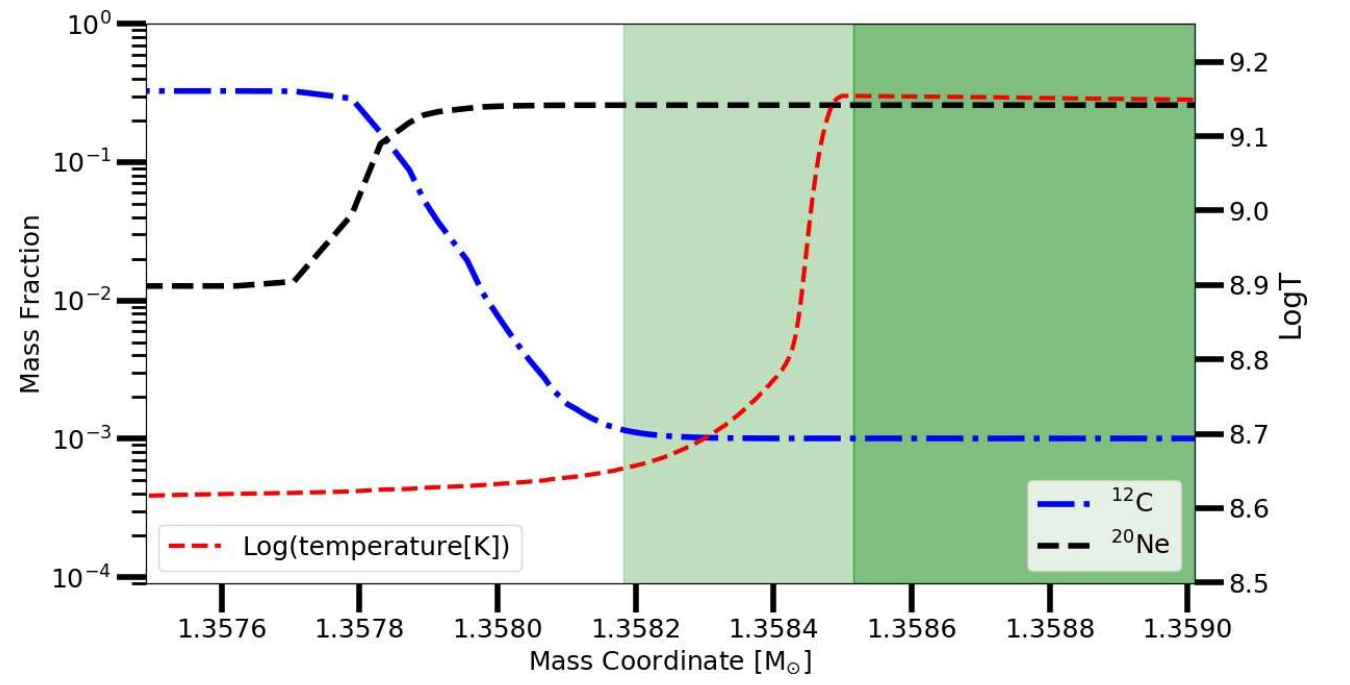

Fig. 6: The temperature, $^{20}$Ne, and $^{12}$C mass fraction profiles as functions of the Lagrangian mass coordinate are shown in the vicinity of the base of the C-flame convective zone in Model C. The dark and light green shaded areas represent the stellar regions dominated by convection and convective boundary mixing, respectively. Across the latter, convective velocities decrease by a factor of 1000.

Figure 6 shows the temperature, $^{20}$Ne, and $^{12}$C mass fraction profiles in the vicinity of the bottom of the C-flame convective zone. As previously discussed, the location of the bottom of the C-shell convective zone during the first C-flash remains tightly bound to that of the temperature maximum. However, convective boundary mixing (CBM) penetrates into the convectively stable layers below the convective boundary and homogenizes the $^{12}$C abundance distribution, mixing it with the $^{12}$C-depleted material from the C-burning convective zone. This results in a low and nearly flat $^{12}$C abundance in the stable layers below the convective boundary. Consequently, the steep rise in $^{12}$C is shifted by $\sim 3\times10^{-4}\,M_\odot$ away from the location of the peak temperature. As a result, a decrease in $\log T$ immediately below this point cannot be compensated by a sufficiently strong increase in the $^{12}$C mass fraction. Under these conditions, C-flame propagation becomes significantly more difficult than in the absence of CBM, until the C abundance behind the C-flame is reduced to such a low level that its further ignition becomes impossible (see also Denissenkov et al. 2013). At that point, the flame will be quenched before reaching the WD center.

Timmes et al. (1994) pointed out that a physically consistent simulation of the C-flame requires a very fine spatial resolution, with mass zones thinner than $\sim$1km in the burning region. In our simulations of C-flame propagation, we include nearly 50 mass zones within 1km around the temperature peak in the C-burning region, as shown in Figure 7.

## 4. Trans-Fe nucleosynthesis

During C-burning, the stellar plasma in our models is highly enriched in $\alpha$-particles. This is due to their release via the $^{12}$C($^{12}$C,$\alpha$)$^{20}$Ne reaction. These $\alpha$-particles are then captured by $^{22}$Ne and converted into neutrons via the $^{22}$Ne($\alpha$,n)$^{25}$Mg reaction.

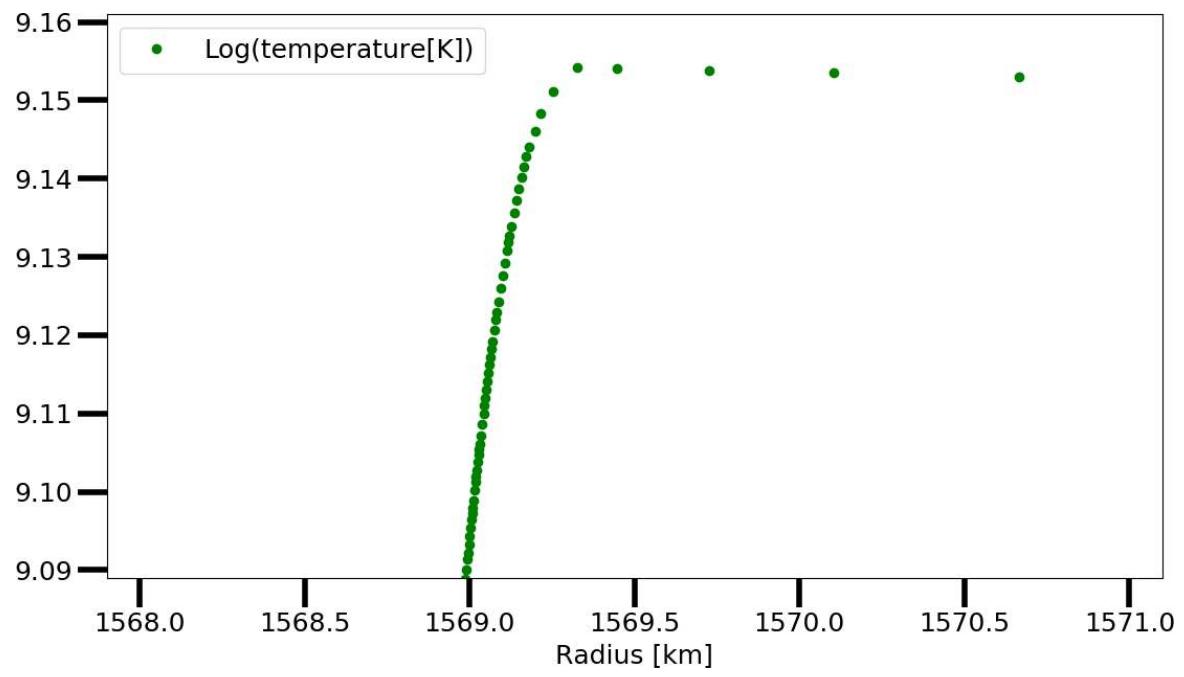

Fig. 7: A typical mass zoning in the region of the temperature peak within the C-flame front in our simulations.

Table 3: Mean neutron densities during surface C-burning, duration of the $1^{st}$ C-flash and total number of surface C-flashes for all the models presented in this work.

| Model | Mean n-dens | Δt | # |
|---|---|---|---|
| | $1^{st}$ C-flash [cm$^{-3}$] | $1^{st}$ C-flash [yr] | C-flashes |
| A | 6.9 $10^{14}$ | 146 | 1 |
| B | 2.3 $10^{14}$ | 153 | 1 |
| C | 5.4 $10^{13}$ | 252 | 3 |
| D | 1.5 $10^{13}$ | 553 | 15 |
| E | 4.2 $10^{13}$ | 233 | 2 |

The rate of neutron release establishes a high neutron density, ranging between $\sim10^{13}$ and $\sim10^{14}$ neutrons cm$^{-3}$ during surface C-burning. Additionally, the earlier surface C-burning ignition in Model D (in terms of WD mass) allows mass accretion to continue for a longer period after the first C-flash, before approaching the Chandrasekhar limit and igniting C-burning explosively in the center. This is why Model D experiences a much higher total number of surface C-flashes compared to the other models (see Table 3).

The abundance profiles of selected species at the top of the primary WD at the onset of explosive C-burning in the center are shown in Figure 8. We identify the onset of explosive central carbon ignition as the epoch when the nuclear burning timescale becomes comparable to the local convective timescale at the center. In our models, this happens at a central temperature of approximately $7.9{\times}10^8$K, in excellent agreement with Woosley et al. (2004), who, guided by multi-dimensional hydrodynamic simulations, identified the onset of the runaway at average central temperatures between 7.7 and $7.9{\times}10^8$K. This marks the simmering phase and the development of a convective core prior to the final explosion. In all our models, trans-Fe elements are synthesized on the WD surface, forming a distribution peaked at Kr. For example, $^{86}$Kr has a mass fraction ranging between $10^{-5}$ and $3{\times}10^{-5}$, corresponding to an overproduction factor between $\sim$500 and $\sim$1300 compared to solar. The surface layer enriched in heavy elements spans a thickness in mass between $\sim$0.04 and $\sim$0.11 $M_\odot$. The bottom boundary of this trans-Fe-rich external layer is located at a mass coordinate close to the primary WD mass at the moment of surface C-burning ignition, while the top boundary lies close to the surface, at a near-Chandrasekhar mass coordinate. Since Model D ignited surface C-burning earlier and at a lower WD mass than all other models, it features the thickest trans-Fe-enriched layer ($\sim$0.11 $M_\odot$). Integrating the Kr isotope abundances over the entire trans-Fe layer yields a total Kr mass of $\sim 4{\times}10^{-6}$ $M_\odot$ produced in Model D.

Moreover, each of the 15 surface C-flashes in Model D impacted the $^{12}$C and $^{22}$Ne abundance profiles, producing abundance dips at every zone where C-burning and $^{22}$Ne+$\alpha$ reactions were activated. A similar effect is also visible in other models, e.g., Model C, which experiences three surface C-flashes.

Figure 9 shows the surface structure evolution and the development of five of the 15 C-flash episodes in Model D. Each C-flash episode drives the formation of a convective zone, typically between $5{\times}10^{-3}$ $M_\odot$ and $10^{-2}$ $M_\odot$ in size, in which trans-Fe elements are synthesized via neutron captures. We show key species and temperature profiles at the onset of two of these five C-flashes in Figure 10. The temperature peaks at $\sim$1.3GK at the base of each C-flash-driven convective zone. Its location coincides with the step in the abundance profiles of both $^{20}$Ne and $^{86}$Kr, indicating active C-burning and trans-Fe element production, respectively.

All models show strong production of $^{28}$Si via the capture of $\alpha$ particles on the abundant $^{20}$Ne produced during C-burning. $^{28}$Si is overproduced by a factor of $\sim$1000 compared to solar in all models except Model D, where it is overproduced by a factor of only $\sim$10. This difference is due to the lower surface C-burning temperature in Model D, caused by the lower primary WD mass at the time of ignition.

In Figure A.1, the final isotopic abundance distribution in the external trans-Fe-enriched layers of the models listed in Table 2 is shown. All abundances have been extracted at the same mass coordinate, $M_r = 1.35\ M_\odot$, which is representative of trans-Fe element production in all models (see Figure 8). The nucleosynthetic production in all models peaks at $^{84}$Kr and $^{86}$Kr. The highest production of these isotopes is reached in models B and D, with $^{84}$Kr overproduced by a factor of $\sim$540 (in Model B) and $^{86}$Kr by a factor of $\sim$1300 (in both models B and D) compared to solar. The nucleosynthetic production in all models extends beyond the first-peak elements, reaching species around atomic mass number $A = 120$. The heaviest species significantly overproduced compared to solar (by a factor of 30) in models B and E is $^{116}$Sn, while in models C and D it is $^{123}$Sb. Additionally, up to $\sim 6{\times}10^{-6}$ $M_\odot$ of $^{60}$Fe is produced in these same outer layers.

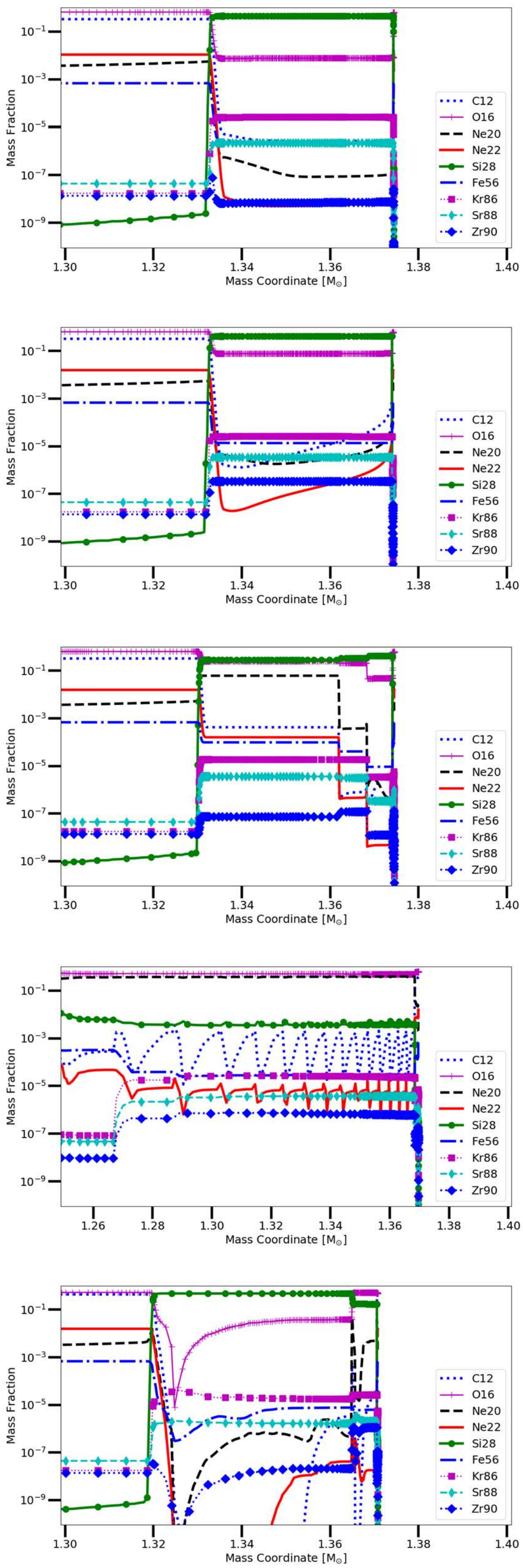

Fig. 8: Trans-Fe elements abundance profile in the external layers of the near-Chandrasekhar primary WD for the models listed in Table 2. Models from A to E are shown from top to bottom.

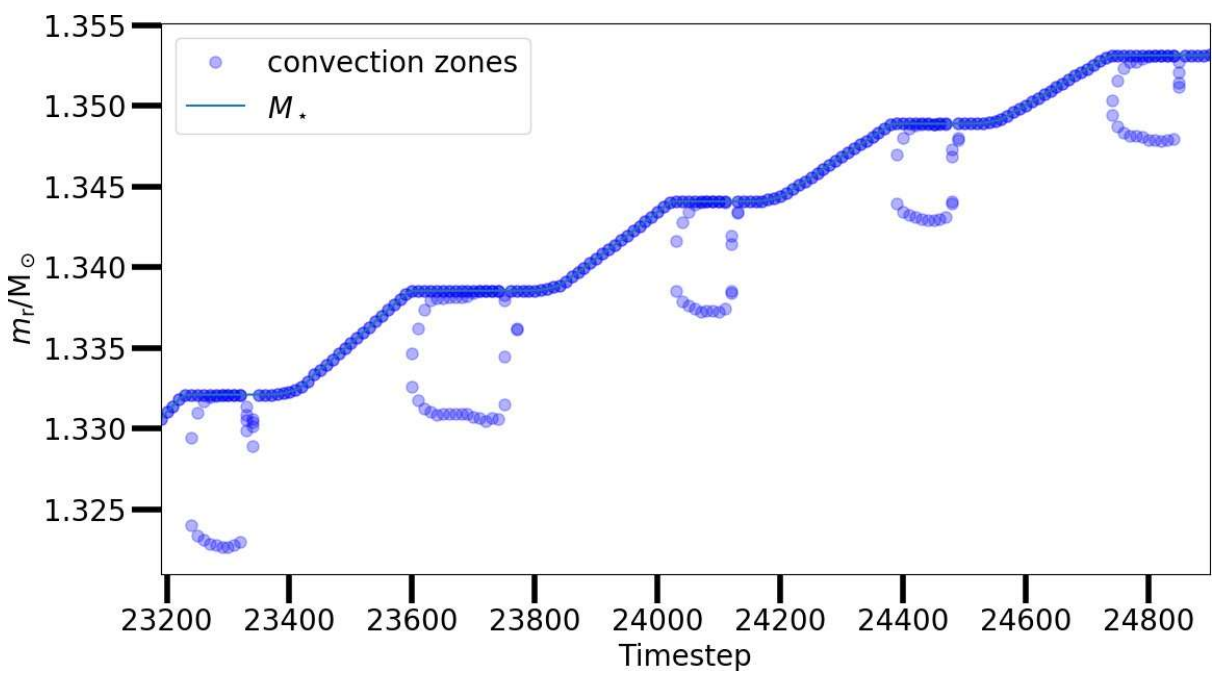

Fig. 9: Kippenhahn diagram showing the surface structure evolution and the development of five C-flashes episodes in Model D. The time coordinate on the X-axis is expressed in computational timesteps to ensure a proper graphical resolution of the C-flashes-driven convective zones.

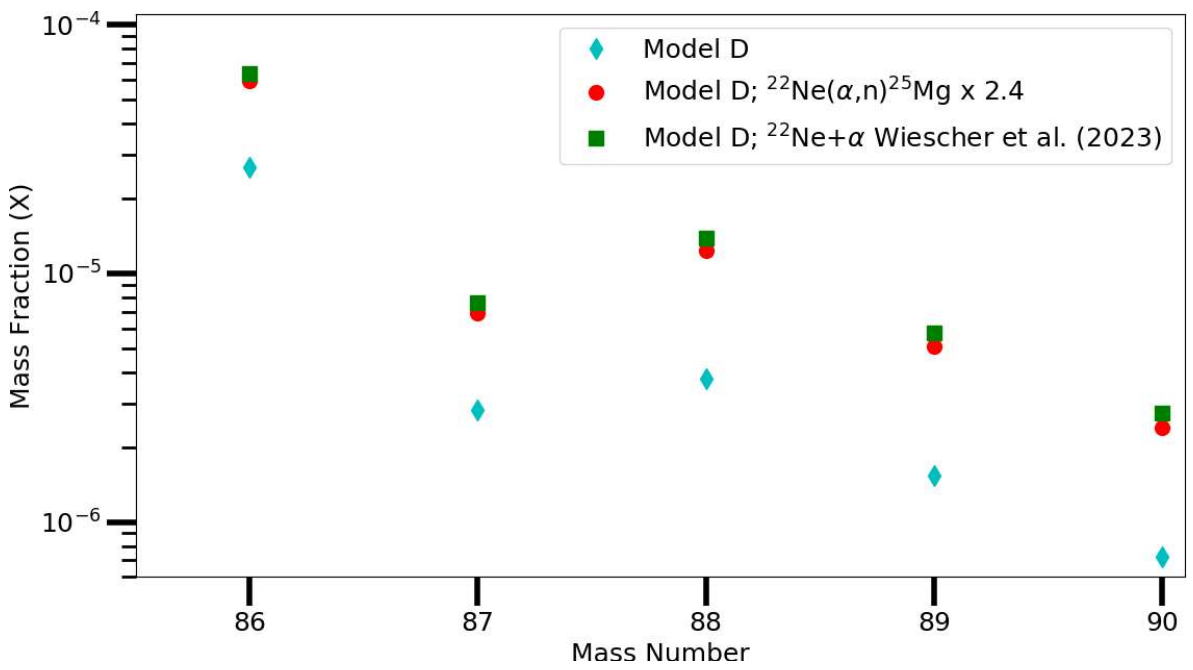

Fig. 11: Final abundance in mass fraction of $^{86}$Kr, $^{87}$Rb, $^{88}$Sr, $^{89}$Y and $^{90}$Zr in the external layers of Model D employing different $^{22}Ne(\alpha,n)^{25}Mg$ and $^{22}Ne(\alpha,\gamma)^{26}Mg$ nuclear reaction rates. See main text for details.

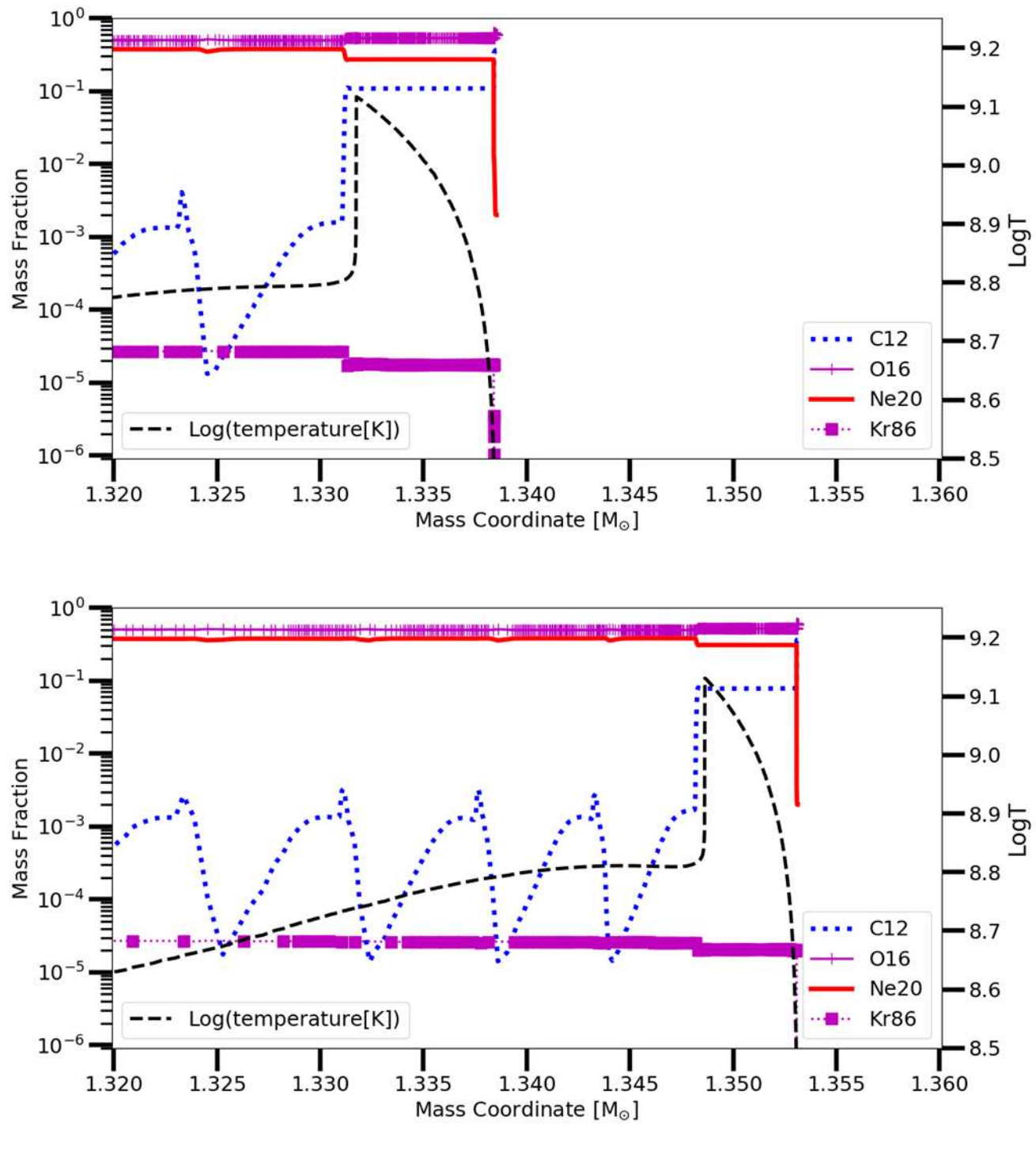

Fig. 10: Key species and temperature profiles at the onset of two of the five C-flashes in Model D shown in figure 9.

We finally performed two additional tests to study the impact of the current uncertainty affecting the $^{22}Ne(\alpha,n)^{25}Mg$ and $^{22}Ne(\alpha,\gamma)^{26}Mg$ reaction rates. In recent years, the reaction rates of both $^{22}Ne(\alpha,n)^{25}Mg$ and $^{22}Ne(\alpha,\gamma)^{26}Mg$ have been re-evaluated by different research groups, leading to contradictory results (see Best et al. 2025, for a review of the latest experimental and theoretical work). For example, Wiescher et al. (2023) showed that their newly evaluated $^{22}Ne(\alpha,n)^{25}Mg$ rate is a factor of three higher at typical He-burning temperatures ($0.2$ GK $< T < 0.3$ GK) and a factor of two lower at temperatures around $T \sim 1$ GK than the rates provided in Adsley et al. (2021), which include results from Ota et al. (2020); Jayatissa et al. (2020), and Ota et al. (2021). The results of our tests are summarised in Figure 11, which shows the final mass fraction abundances of $^{86}$Kr, $^{87}$Rb, $^{88}$Sr, $^{89}$Y, and $^{90}$Zr in the external layers of Model D using different $^{22}Ne(\alpha,n)^{25}Mg$ and $^{22}Ne(\alpha,\gamma)^{26}Mg$ nuclear reaction rates. Switching from the $^{22}Ne(\alpha,n)^{25}Mg$ and $^{22}Ne(\alpha,\gamma)^{26}Mg$ rates of Adsley et al. (2021) (used for all the models in Table 2) to the recently re-evaluated rates of Wiescher et al. (2023) results in an increase of the first-peak isotopes by factors of 2 to 4. This leads to an isotopic overproduction of $^{86}$Kr compared to solar by more than a factor of 3000, and to an elemental overproduction of Kr by a factor of $\sim$800. Interestingly, the results obtained using the reaction rates from Wiescher et al. (2023) closely match those obtained by increasing the $^{22}Ne(\alpha,n)^{25}Mg$ rate by a factor of 2.4 at all temperatures, while keeping the $^{22}Ne(\alpha,\gamma)^{26}Mg$ rate unchanged. This is consistent with a mean temperature during surface C-burning in Model D of $\sim 4.5\times10^8$ K. At that temperature, the $^{22}Ne(\alpha,n)^{25}Mg$ rate from Wiescher et al. (2023) differs from that of Adsley et al. (2021) by the same factor, while the respective $^{22}Ne(\alpha,\gamma)^{26}Mg$ rates differ by only a few percent. Furthermore, this temperature is in agreement with an average between the peak temperature reached during C-flashes, as shown in Figure 10, during which most ($\sim$70%) of the trans-Fe element production takes place, and the $^{22}Ne(\alpha,n)^{25}Mg$ activation in the interpulse period between flashes at a typical temperature of $\simeq 3\times10^8$ K.

Table A.1 shows the mass fraction abundances of selected species among those presented in Figure A.1. These include the abundances in the external layers of the near-Chandrasekhar primary WD for all models listed in Table 2, along with the recomputed Model D using the $^{22}Ne(\alpha,n)^{25}Mg$ and $^{22}Ne(\alpha,\gamma)^{26}Mg$ reaction rates from Wiescher et al. (2023). These abundances can be used as pre-supernova initial conditions for explosive nucle-

osynthesis calculations, applicable to the outermost layers of an exploding near-Chandrasekhar WD.

## 5. Explosive nucleosynthesis

The main goal of this study is to present, for the first time, the full nucleosynthesis, including trans-Fe elements, that occurs during the formation of a near-Chandrasekhar WD via the slow merger of two CO WDs. The primary focus is on the progenitor phase rather than the subsequent explosion. A full grid of explosive nucleosynthesis models will be presented in a follow-up study, which will constitute the second paper in this series. However, in this work, we provide preliminary, yet qualitatively valuable, results by employing a computationally lightweight two-dimensional (2D) SNIa model to obtain an initial estimate of the total ejected yields of trans-Fe elements synthesized during the progenitor phase.

The explosive nucleosynthesis was calculated using tppnp (Tracer Particle Post-Processing Network – Parallel), one of the NuGrid suite of codes described in Jones et al. (2019). Tppnp processes Lagrangian tracer particles from hydrodynamical simulations, where no mixing occurs between particles. Such mixing would alter both the chemical composition and thermodynamic conditions, significantly affecting nucleosynthesis. However, during a Type Ia supernova explosion, material exchange between tracer particles is not expected due to the very short timescale (less than $\sim$1s) taken by the shock front to pass through a given particle. Additionally, Seitenzahl et al. (2010) demonstrated that two-dimensional SNIa simulations with $32^2$ tracer particles reproduce the reference values of the most abundant nuclei ($X(i) \geq 10^{-5}$) globally at the 5 per cent level, with a few exceptions, most notably isotopes of Ne, Mg, and Al. Thus, tppnp's approach, considering a large enough number of particles and treating each one as an independent thermodynamic trajectory with a fixed initial composition and evolving temperature and density conditions, is justified. These conditions, along with the initial composition at the explosion onset, govern nuclear burning within the flame front and the subsequent freeze-out.

The 1.4 $M_\odot$ model of Townsley et al. (2016) (hereafter T16) with reconstructed flame fronts was used as the baseline model in this work. Calculations were carried out using the same method outlined in Keegans et al. (2023). In the T16 model, explosive nuclear combustion of the degenerate carbon–oxygen mixture proceeds in both deflagration and detonation combustion modes, combined into the deflagration–detonation transition (DDT) scenario (Khokhlov 1991). A total of 7856 tracer particles were selected for post-processing nucleosynthesis calculations in the T16 model. Convergence studies on non-reconstructed flame fronts showed that this number of particles is sufficient for converged abundances, as demonstrated in Keegans et al. (2023), and fully adequate for the scope of this first paper. In the T16 model, resolution in the hotter regions of the explosion is very high; however, the resolution in the outer layers of the WD is not as high as that in, e.g., Travaglio et al. (2011). Therefore, there may be some loss of accuracy in the outer layers, where the dominant contribution to the yields of the trans-Fe elements originates. Multi-dimensional models, including $\geq 5\times10^5$ particles and highly resolved external layers, will be considered in the second paper of this series. As described in T16, the explosion in the WD progenitor is initiated through the introduction of an artificial hot spot at a predetermined radius, where the temperature is raised to 10GK. Heat diffuses from this region, igniting the flame in the C/O material.

Our nuclear reaction network used during the explosive nucleosynthesis simulations consists of over 5200 isotopes and 75000 reactions. Reaction rates are sourced from a variety of compilations. The majority of reactions, excluding weak reactions, are taken from Rauscher & Thielemann (2000) (Basel REACLIB), Cyburt et al. (2010) (JINA REACLIB), Dillmann et al. (2006) (KADoNIS), Angulo et al. (1999) (NACRE), Caughlan & Fowler (1988), and Iliadis et al. (2001). Weak reaction rates are compiled from Fuller et al. (1985), Oda et al. (1994), Goriely (1999), and Langanke & Martínez-Pinedo (2000), as well as from Takahashi & Yokoi (1987), as discussed in Jones et al. (2019) and Pignatari et al. (2016).

Table 4 shows the elemental ejected yields beyond Fe-peak elements and up to Ba from our near-Chandrasekhar explosive model. To compute the ejected yields in the third column, the initial abundances adopted at the onset of the explosion are those obtained from our progenitor Model D, with the pre-supernova nucleosynthesis calculated using the $^{22}$Ne($\alpha$,$n$)$^{25}$Mg and $^{22}$Ne($\alpha$,$\gamma$)$^{26}$Mg reaction rates from Wiescher et al. (2023). In particular, we employed the abundances from the corresponding column in Table A.1 in the outermost 0.11 $M_\odot$ layers of our explosive model. In order to facilitate the assessment of the effect of pre-explosive trans-Fe production, the fourth column also provides the final yields from the same explosive model, but without the inclusion of trans-Fe element enrichment in the outer layers of the primary WD, and with a $^{22}$Ne abundance equal to the accreted one for model D (as specified in Table 1). The element ejected in the highest amount is Kr ($6.66\times10^{-6}$ $M_\odot$). ($\gamma$,n) and (n,$\gamma$) reactions during the explosion slightly modify the total abundances synthesized during the progenitor phase (see also Arnould (1976); Travaglio et al. (2011); Roberti et al. (2023)). The total ejected amount of all first-peak elements is reduced by $\sim$30% compared to what was produced during pre-supernova nucleosynthesis. The only exception is Sr, whose abundance increases by $\sim$17% in the explosive ejected yields. Most importantly, Kr, Rb, and Sr are all ejected in significant amounts ($\geq 10^{-6}$ $M_\odot$) during the explosion. This is comparable with the abundances reported by Watson et al. (2019), who detected $\sim 10^{-5}$ $M_\odot$ of Sr from near-IR and partially near-UV lines in the AT2017gfo kilonova spectra (see their Figure 4). This suggests that the identification of trans-Fe elements, formed exclusively during the progenitor phase under near-Chandrasekhar conditions, is feasible, and could potentially serve as a novel "smoking gun" for distinguishing near-Chandrasekhar explosions from sub-Chandrasekhar ones. However, as already noted, these encouraging results require confirmation from a grid of higher-resolution SNIa models, which will be presented in a follow-up study.

Finally, we report a total ejected $^{60}$Fe mass of $\sim 4.6\times10^{-7}$ $M_\odot$. This represents an increase by a factor of $\sim$20 in the $^{60}$Fe yields compared to those obtained from the same explosive model without the inclusion of trans-Fe element enrichment in the external layers of the primary WD. This might impact the theoretical mass estimate of $^{60}$Fe in the Milky Way depending on the rate of near-Chandrasekhar explosions, since SNIa are generally thought to contribute negligibly to its content in the Galaxy (Vasini et al. 2022). Additionally, this could be a promising case for next-generation $\gamma$-ray telescopes, which will have the sensitivity to directly detect a single supernova remnant (Scharrer et al. 2025).

Table 4: Ejected Yields in $M_\odot$ of trans-Fe elements from the T16 explosion model adopting our final pre-explosive surface trans-Fe abundances as initial abundances.

| Element | Atomic number | Ejected mass [$M_\odot$] | Ejected mass [$M_\odot$] No initial trans-Fe |
|---|---|---|---|
| Ga | 31 | 1.36e-06 | 7.89e-09 |
| Ge | 32 | 6.03e-06 | 1.93e-07 |
| As | 33 | 8.10e-07 | 2.79e-09 |
| Se | 34 | 5.00e-06 | 2.53e-07 |
| Br | 35 | 4.37e-07 | 1.16e-11 |
| Kr | 36 | 6.66e-06 | 1.61e-09 |
| Rb | 37 | 1.41e-06 | 4.29e-13 |
| Sr | 38 | 1.74e-06 | 2.21e-11 |
| Y | 39 | 3.70e-07 | 1.23e-14 |
| Zr | 40 | 7.64e-07 | 2.33e-13 |
| Nb | 41 | 1.63e-09 | 3.75e-17 |
| Mo | 42 | 5.10e-08 | 2.34e-16 |
| Tc | 43 | 5.22e-10 | 6.77e-19 |
| Ru | 44 | 7.29e-09 | 3.76e-17 |
| Rh | 45 | 5.20e-10 | 2.31e-19 |
| Pd | 46 | 1.94e-08 | 2.98e-18 |
| Ag | 47 | 1.48e-10 | 9.78e-24 |
| Cd | 48 | 1.95e-08 | 7.53e-24 |
| In | 49 | 1.03e-10 | 4.31e-30 |
| Sn | 50 | 5.56e-08 | 1.08e-31 |
| Sb | 51 | 1.41e-09 | 2.82e-39 |
| Te | 52 | 2.94e-09 | 1.30e-41 |
| I | 53 | 6.48e-10 | 9.31e-45 |
| Xe | 54 | 1.20e-08 | 2.01e-47 |
| Cs | 55 | 1.16e-09 | 1.68e-50 |
| Ba | 56 | 1.62e-08 | 2.57e-53 |
| [Rb/Sr] | | 0.27 | |
| [Rb/Zr] | | 0.34 | |

**Notes.** Surface trans-Fe abundances are from our Model D progenitor and are applied in the outermost $\sim$0.11 $M_\odot$. The adopted $^{22}$Ne($\alpha$,$n$)$^{25}$Mg and $^{22}$Ne($\alpha$,$\gamma$)$^{26}$Mg reaction rates are from Wiescher et al. (2023). The last column shows the final yields from the same explosive model, but without the inclusion of trans-Fe element enrichment in the external layers of the primary WD, and with a $^{22}$Ne abundance equal to the accreted one for model D (as specified in table 1). The elemental ratios [Rb/Sr] and [Rb/Zr] are also given in square bracket notation in the last two lines.

## 6. Conclusions

In this work, we calculated the evolution and nucleosynthesis in slowly merging CO WDs using the stellar code MESA. We presented a grid of five models detailing the evolution and nucleosynthesis of slowly merging CO white dwarfs approaching the Chandrasekhar mass. Our main results and final remarks can be summarised as follows:

1. As our models approach the Chandrasekhar mass during the merger phase, carbon-burning ignites on the surface before it does in the center. The stellar plasma in our models gets enriched in $\alpha$-particles, due to their release via the $^{12}$C($^{12}$C,$\alpha$)$^{20}$Ne reaction. These $\alpha$-particles are then captured by $^{22}$Ne and converted into neutrons via the $^{22}$Ne($\alpha$,$n$)$^{25}$Mg reaction in the external layers of the primary WD.
2. The resulting neutron-capture seed distribution closely resembles a weak $s$-process one and peaks at Kr, which is overproduced by a factor of $\sim$800 compared to solar, with specific isotopes (in particular $^{86}$Kr) being overproduced by more than a factor of 3000. Current uncertainties affecting the $^{22}$Ne($\alpha$,$n$)$^{25}$Mg and $^{22}$Ne($\alpha$,$\gamma$)$^{26}$Mg reaction rates impact these results by about a factor of 2. The $^{22}$Ne($\alpha$,$n$)$^{25}$Mg and $^{22}$Ne($\alpha$,$\gamma$)$^{26}$Mg nuclear reaction rates tested in the stellar models presented in this work are available in tabulated format from the open-source platform ChANUREPS (ChETEC AstroNUclear REPositorieS [2].
3. The mass of the outermost layers enriched in first-peak $s$-process elements depends strongly on the accretion rate onto the primary WD and ranges between $\sim$0.04 $M_\odot$ and $\sim$0.11 $M_\odot$.
4. We also computed the nucleosynthesis arising from the explosion of a near-Chandrasekhar explosion model when the final abundances obtained from our progenitors are adopted as the initial conditions for the explosion simulation. Ga, Ge, Se, Kr, Rb, and Sr are all ejected in high amounts ($>10^{-6}$ $M_\odot$) during the explosion. In Battino et al. (2020), we found that a similar nucleosynthesis process during the progenitor phase may also occur on the surface of near-Chandrasekhar white dwarfs formed through the accretion of H-rich material via the single-degenerate scenario.
5. These ejected abundances are comparable to what was reported by Watson et al. (2019), who successfully detected $\sim10^{-5}$ $M_\odot$ of Sr from near-IR and partially near-UV lines of the AT2017gfo kilonova spectra. This suggests that spectral identification of trans-Fe elements in SNIa is not unrealistic and could constitute a new spectroscopically observable "smoking gun" of a near-Chandrasekhar progenitor. Crucially, and to the best of our knowledge, the synthesis of trans-Fe elements during the SNIa progenitor phase is only possible if the exploding WD has a mass close to the Chandrasekhar limit. These encouraging results need to be confirmed by a grid of more highly resolved SNIa models, which we will present in a follow-up study.
6. On the other hand, a non-detection of trans-Fe elements in the SN ejecta would not necessarily imply that the explosion originated from a sub-Chandrasekhar progenitor. For instance, an accretion rate $< 2\times10^{-6}$M$_\odot$ would produce a Chandrasekhar-mass primary WD without igniting C on the surface, and therefore without producing trans-Fe elements. This means that the number of nearby SNIa showing a trans-Fe element signature in their ejecta could instead provide a hard lower limit on the near-Chandrasekhar fraction of the total SNIa population, thereby setting an important and robust constraint.
7. The only sub-Chandrasekhar progenitor scenario that might still allow some production of trans-Fe elements while also predicting light curves and spectra consistent with those of normal SNIa is the accretion of He-rich matter onto a massive CO white dwarf, provided that the CO core is sufficiently massive ($\geq$1M$_\odot$) and the He buffer is small enough ($\leq$0.05M$_\odot$). Otherwise, a low-luminosity and peculiar nucleosynthesis event will result (Woosley & Kasen 2011). In this scenario, trans-Fe nucleosynthesis might occur if a specific variable accretion rate is realized, consisting of an initial high accretion rate ($\sim10^{-6}$ M$_\odot$ yr$^{-1}$) to ensure steady accretion of the primary CO WD beyond $\sim$1M$_\odot$, followed by a lower accretion rate phase ($\sim10^{-7}$ M$_\odot$ yr$^{-1}$) sustained long enough to allow the development of strong He flashes and activation of the $^{22}$Ne($\alpha$,$n$)$^{25}$Mg reaction, and finally a very low accretion rate phase ($\sim10^{-8}$ M$_\odot$ yr$^{-1}$) to trigger the final detonation

[2] http://chanureps.chetec-infra.eu/)

(Piersanti et al. 2014). While the simultaneous fulfillment of all these conditions may challenge the efficiency of this scenario, it still represents an unexplored case in terms of the resulting nucleosynthesis. Addressing this will be essential for developing a comprehensive understanding of all scenarios capable of producing a trans-Fe spectroscopic signature, and will be the main goal of a future work in this paper series.

8. Finally, we report a total ejected $^{60}$Fe mass of $\sim 4.6\times10^{-7}$ $M_{\odot}$. This corresponds to an increase by a factor of ∼20 in the $^{60}$Fe yields compared to those obtained from the same explosive model without the inclusion of trans-Fe element enrichment in the external layers of the primary WD and, depending on the rate of near-Chandrasekhar explosions, might impact the theoretical mass estimate of $^{60}$Fe in the Milky Way. Moreover, this could be a promising case for next-generation γ-ray telescopes, which will have the sensitivity to directly detect a single supernova remnant.

The validation of these predictions heavily relies on upcoming large observational surveys, such as the Time-Domain Extragalactic Survey (TiDES). TiDES will spectroscopically follow up transients detected by the VRO-LSST survey and will observe spectra of over 30,000 supernovae, many of which will be SNIa. This strongly motivates the computation of a full grid of highly resolved multi-D explosion simulations incorporating the trans-Fe enrichment from our progenitor models, already underway and to be published in a follow-up study. This will ultimately pave the way for the computation of synthetic observables of trans-Fe elements in SNIa ejecta, and hence for the search for their spectroscopic signatures in SNIa emissions.

*Acknowledgements.* This article is based upon work from the ChETEC COST Action (CA16117). UB acknowledges support by ChETEC-INFRA (EU project no. 101008324), JINA-CEE (NSF Grant PHY-1430152), the European Research Council (ERC-2019-STG Nr. 852016) and the Italian Ministry of Research (FARE project EASγ R20SLAA8CJ ). This research was enabled in part by support provided by the BC DRI Group, the Digital Research Alliance of Canada (alliance.can.ca). and ongoing access to viper, the University of Hull High Performance Computing Facility. We thank the IReNA network supported by US NSF AccelNet. We acknowledge the EuroHPC Joint Undertaking for awarding this project access to the EuroHPC supercomputer LUMI, hosted by CSC (Finland) and the LUMI consortium through a EuroHPC Benchmark Access call (proposal ID: 465000917). The work of FKR is supported by the Klaus Tschira Foundation, by the Deutsche Forschungsgemeinschaft (DFG, German Research Foundation) – RO 3676/7-1, project number 537700965, and by the European Union (ERC, ExCEED, project number 101096243). Views and opinions expressed are, however, those of the authors only and do not necessarily reflect those of the European Union or the European Research Council Executive Agency. Neither the European Union nor the granting authority can be held responsible for them. This work benefited from discussions and resources provided by the STFC-funded BRIDGCE UK Network (https://www.bridgce.ac.uk/). SAS acknowledges funding from STFC grant ST/X00094X/1.

## Appendix A: Additional material

Table A.1: Pre-supernova abundances in mass fraction of selected species among those presented in Fig. A.1.

| Isotope | Model A | Model B | Model C | Model D | Model D $^{22}Ne+\alpha$ Wiescher et al. (2023) | Model E |
|---|---|---|---|---|---|---|
| $^{16}O$ | 7.62e-03 | 7.96e-02 | 2.54e-01 | 4.95e-01 | 4.89e-01 | 3.63e-02 |
| $^{20}Ne$ | 1.02e-07 | 1.87e-06 | 6.21e-02 | 3.84e-01 | 3.87e-01 | 5.88e-07 |
| $^{22}Ne$ | 5.04e-09 | 8.75e-08 | 1.59e-04 | 5.60e-06 | 2.81e-05 | 2.29e-09 |
| $^{23}Na$ | 8.65e-08 | 1.26e-07 | 3.12e-03 | 8.12e-03 | 8.01e-03 | 1.52e-07 |
| $^{24}Mg$ | 1.76e-05 | 2.62e-04 | 2.22e-02 | 6.10e-02 | 6.23e-02 | 6.86e-05 |
| $^{25}Mg$ | 1.11e-06 | 6.89e-06 | 5.46e-03 | 1.39e-02 | 1.34e-02 | 7.52e-07 |
| $^{26}Mg$ | 4.78e-02 | 4.55e-02 | 3.04e-02 | 2.94e-02 | 2.97e-02 | 5.69e-02 |
| $^{27}Al$ | 3.12e-03 | 1.28e-02 | 2.74e-02 | 5.00e-03 | 5.11e-03 | 5.76e-03 |
| $^{28}Si$ | 4.43e-01 | 4.24e-01 | 2.89e-01 | 3.80e-03 | 3.86e-03 | 4.83e-01 |
| $^{32}S$ | 3.96e-01 | 3.52e-01 | 2.46e-01 | 1.34e-06 | 1.38e-06 | 3.52e-01 |
| $^{52}Cr$ | 3.51e-08 | 8.02e-08 | 1.23e-06 | 5.37e-07 | 3.94e-07 | 7.08e-08 |
| $^{53}Cr$ | 2.20e-08 | 2.32e-08 | 3.55e-07 | 2.12e-07 | 1.54e-07 | 1.77e-08 |
| $^{54}Cr$ | 5.04e-06 | 1.65e-06 | 1.61e-06 | 2.20e-06 | 1.72e-06 | 3.26e-06 |
| $^{55}Mn$ | 2.37e-07 | 2.98e-07 | 2.80e-06 | 1.02e-06 | 7.89e-07 | 2.46e-07 |
| $^{56}Fe$ | 2.10e-06 | 1.36e-05 | 9.74e-05 | 3.07e-05 | 2.17e-05 | 7.10e-06 |
| $^{57}Fe$ | 6.92e-07 | 2.09e-06 | 2.78e-05 | 1.49e-05 | 1.02e-05 | 6.77e-07 |
| $^{58}Fe$ | 3.35e-04 | 2.93e-04 | 1.10e-04 | 7.54e-05 | 5.37e-05 | 3.34e-04 |
| $^{60}Fe$ | 1.47e-06 | 1.92e-07 | 3.27e-05 | 5.63e-05 | 3.96e-05 | 2.20e-07 |
| $^{59}Co$ | 4.98e-07 | 1.51e-06 | 9.88e-05 | 1.02e-04 | 7.37e-05 | 1.05e-06 |
| $^{60}Ni$ | 5.82e-08 | 7.03e-07 | 1.99e-05 | 5.73e-05 | 4.07e-05 | 5.66e-07 |
| $^{61}Ni$ | 4.52e-08 | 2.45e-07 | 1.58e-05 | 1.91e-05 | 1.45e-05 | 1.25e-07 |
| $^{62}Ni$ | 2.96e-04 | 4.14e-04 | 1.49e-04 | 1.00e-04 | 8.87e-05 | 4.14e-04 |
| $^{63}Ni$ | 1.86e-06 | 1.10e-06 | 4.16e-05 | 7.85e-05 | 7.16e-05 | 4.42e-08 |
| $^{64}Ni$ | 1.86e-04 | 3.76e-05 | 7.45e-05 | 1.22e-04 | 1.28e-04 | 7.23e-05 |
| $^{65}Cu$ | 2.49e-07 | 1.96e-07 | 1.25e-05 | 2.63e-05 | 2.85e-05 | 3.65e-07 |
| $^{66}Zn$ | 3.12e-08 | 2.42e-06 | 1.62e-05 | 2.69e-05 | 3.02e-05 | 2.60e-07 |
| $^{67}Zn$ | 7.28e-10 | 2.42e-08 | 6.03e-06 | 1.24e-05 | 1.42e-05 | 1.71e-10 |
| $^{68}Zn$ | 5.99e-07 | 7.48e-06 | 2.82e-05 | 4.14e-05 | 5.23e-05 | 4.87e-07 |
| $^{70}Zn$ | 1.12e-08 | 2.02e-08 | 1.21e-06 | 1.27e-05 | 1.68e-05 | 2.28e-09 |
| $^{71}Ga$ | 4.11e-10 | 2.71e-09 | 2.25e-06 | 6.30e-06 | 8.38e-06 | 4.69e-10 |
| $^{70}Ge$ | 4.59e-10 | 9.12e-08 | 5.86e-06 | 7.01e-06 | 9.32e-06 | 5.26e-09 |
| $^{72}Ge$ | 8.60e-07 | 1.98e-05 | 8.68e-06 | 6.62e-06 | 9.01e-06 | 1.99e-06 |
| $^{73}Ge$ | 1.24e-08 | 1.20e-07 | 4.75e-06 | 8.36e-06 | 1.12e-05 | 3.16e-10 |
| $^{74}Ge$ | 5.63e-06 | 1.91e-05 | 1.15e-05 | 1.27e-05 | 1.79e-05 | 3.67e-06 |
| $^{76}Ge$ | 2.97e-07 | 1.46e-07 | 9.65e-06 | 8.25e-06 | 1.26e-05 | 5.53e-08 |
| $^{75}As$ | 1.44e-09 | 1.67e-08 | 1.81e-06 | 3.03e-06 | 4.22e-06 | 7.70e-08 |
| $^{76}Se$ | 6.96e-09 | 2.53e-07 | 8.74e-07 | 2.01e-06 | 2.96e-06 | 4.85e-08 |
| $^{77}Se$ | 1.32e-09 | 2.07e-08 | 1.72e-06 | 8.94e-06 | 1.34e-05 | 6.28e-10 |
| $^{78}Se$ | 3.12e-06 | 1.37e-05 | 8.77e-06 | 6.10e-06 | 1.00e-05 | 5.62e-06 |
| $^{79}Se$ | 6.86e-08 | 1.29e-07 | 3.16e-06 | 2.95e-06 | 4.65e-06 | 2.63e-09 |
| $^{80}Se$ | 1.05e-05 | 6.72e-06 | 1.01e-05 | 1.04e-05 | 1.74e-05 | 4.59e-06 |
| $^{82}Se$ | 6.25e-07 | 5.96e-08 | 7.97e-06 | 6.54e-06 | 1.43e-05 | 1.82e-07 |
| $^{81}Br$ | 1.82e-08 | 5.90e-08 | 1.53e-06 | 1.78e-06 | 2.98e-06 | 5.64e-07 |
| $^{82}Kr$ | 6.83e-09 | 4.42e-07 | 6.68e-07 | 7.91e-07 | 1.38e-06 | 2.35e-07 |
| $^{83}Kr$ | 2.30e-09 | 6.35e-08 | 2.05e-06 | 3.37e-06 | 5.99e-06 | 1.06e-08 |
| $^{84}Kr$ | 5.37e-06 | 3.57e-05 | 1.31e-05 | 1.03e-05 | 2.00e-05 | 1.31e-05 |
| $^{86}Kr$ | 2.59e-05 | 2.61e-05 | 1.90e-05 | 2.52e-05 | 6.22e-05 | 1.86e-05 |
| $^{85}Rb$ | 3.71e-10 | 3.05e-09 | 6.30e-07 | 6.08e-06 | 1.21e-05 | 2.61e-08 |
| $^{87}Rb$ | 3.44e-06 | 2.44e-06 | 3.32e-06 | 2.74e-06 | 7.67e-06 | 3.02e-06 |
| $^{88}Sr$ | 2.27e-06 | 3.52e-06 | 3.62e-06 | 3.82e-06 | 1.40e-05 | 1.68e-06 |
| $^{89}Y$ | 4.20e-10 | 5.03e-09 | 2.15e-07 | 1.46e-06 | 5.48e-06 | 4.79e-10 |
| $^{90}Zr$ | 7.06e-09 | 3.35e-07 | 7.59e-08 | 6.65e-07 | 2.51e-06 | 2.01e-08 |

Table A.1: continued.

| Isotope | Model A | Model B | Model C | Model D | Model D $^{22}Ne+\alpha$ Wiescher et al. (2023) | Model E |
|---|---|---|---|---|---|---|
| $^{92}Zr$ | 2.33e-10 | 1.76e-09 | 1.11e-07 | 1.75e-07 | 7.56e-07 | 2.02e-10 |
| $^{100}Ru$ | 5.06e-10 | 9.02e-10 | 1.12e-09 | 9.01e-10 | 5.59e-09 | 4.54e-10 |
| $^{104}Pd$ | 6.56e-10 | 3.28e-09 | 1.94e-10 | 1.71e-10 | 1.38e-09 | 1.30e-09 |
| $^{106}Pd$ | 1.11e-09 | 7.66e-10 | 2.98e-09 | 1.73e-08 | 1.27e-07 | 7.49e-10 |
| $^{110}Cd$ | 4.44e-09 | 7.05e-09 | 7.34e-10 | 1.29e-09 | 1.12e-08 | 5.59e-09 |
| $^{112}Cd$ | 1.76e-09 | 4.11e-10 | 2.09e-09 | 1.28e-08 | 9.78e-08 | 6.37e-10 |
| $^{116}Sn$ | 4.91e-08 | 4.18e-08 | 9.94e-10 | 4.80e-10 | 4.93e-09 | 5.50e-08 |
| $^{118}Sn$ | 1.13e-08 | 1.44e-09 | 5.04e-09 | 9.15e-09 | 7.84e-08 | 3.43e-09 |
| $^{123}Sb$ | 1.84e-15 | 1.63e-14 | 6.81e-10 | 4.51e-09 | 4.38e-08 | 6.83e-16 |
| $^{139}La$ | 3.21e-10 | 1.38e-09 | 1.62e-09 | 1.58e-09 | 3.30e-08 | 1.21e-09 |
| [Rb/Sr] | 0.54 | 0.20 | 0.36 | 0.72 | 0.51 | 0.62 |
| [Rb/Zr] | 2.74 | 0.93 | 1.14 | 0.90 | 0.64 | 2.25 |

**Notes.** Elemental ratios [Rb/Sr] and [Rb/Zr] are also given in the last two lines, adopting the standard square bracket notation: $[X/Y] = \log_{10}((X_*/Y_*)/(X_\odot/Y_\odot))$.

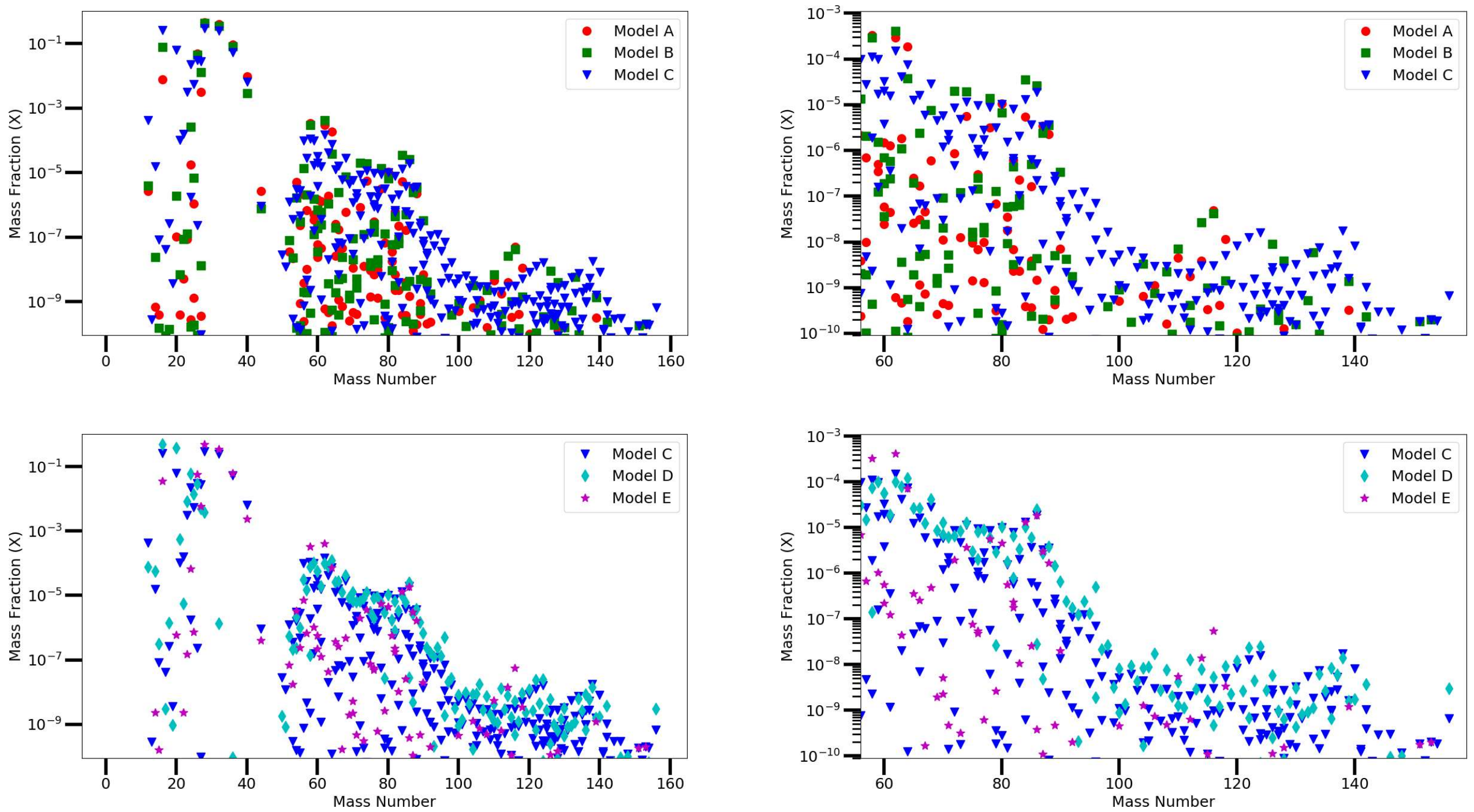

Fig. A.1: Final abundance distribution in the external layers of the near-Chandrasekhar primary WD for the models listed in table 2. All abundances have been extracted at the same mass coordinate, $M_r$ = 1.35 $M_{\odot}$, which is representative of trans-Fe element production in all models. For all models, we present both the distribution over the whole atomic mass range and a zoom in the species beyond Fe (A=56).